\documentclass[pmlr,twocolumn,a4paper,10pt]{jmlr}
\usepackage{isipta2021}
\usepackage[british]{babel}
\usepackage[T1]{fontenc}
\usepackage{color,bm,comment,enumerate,booktabs,tikz,pgfplots,hyperref} 
\pgfplotsset{compat=1.17}
\usetikzlibrary{automata, positioning, arrows,shapes}
 
\jmlrworkshop{ISIPTA 2021}
\title{Robust Model Checking with Imprecise Markov Reward Models}
\author{\Name{Alberto Termine}\Email{alberto.termine@unimi.it}\\
\addr{University of Milan, Milan, Italy}
\AND
\Name{Alessandro Antonucci}\Email{alessandro@idsia.ch}\\
\Name{Alessandro Facchini}\thanks{Supported by the Hasler foundation grant n. 20061.}
\Email{alessandro.facchini@idsia.ch}\\
\addr{Istituto Dalle Molle di Studi sull'Intelligenza Artificiale (IDSIA) - Lugano, Switzerland}
\AND
\Name{Giuseppe Primiero}\Email{giuseppe.primiero@unimi.it}\\
\addr{University of Milan, Milan, Italy}}
\begin{document}
\maketitle
\begin{abstract}
In recent years probabilistic model checking has become an important area of research because of the diffusion of computational systems of stochastic nature. Despite its great success, standard probabilistic model checking suffers the limitation of requiring a sharp specification of the probabilities governing the model behaviour. The theory of imprecise probabilities offers a natural approach to overcome such limitation by a sensitivity analysis with respect to the values of these parameters. However, only extensions based on discrete-time imprecise Markov chains have been considered so far for such a robust approach to model checking. We present a further extension based on imprecise Markov reward models. In particular, we derive efficient algorithms to compute lower and upper bounds of the expected cumulative reward and probabilistic bounded rewards based on existing results for imprecise Markov chains. These ideas are tested on a real case study involving the spend-down costs of geriatric medicine departments.
\end{abstract}
\section{Introduction}\label{sec:intro}
\emph{Model checking} is a fully-automatic logic-based technique to decide whether a model satisfies some given requirements. As such, it has been mostly applied as a tool to decide correctness and compliance in systems design. Relevant examples are software verification \citep[e.g.,][]{MCSoftwareVerification}, system communication and also computational biology \citep[e.g.,][]{MCBiologicalSystems1, MCBiologicalSystems2}. During the last decades, the spread of stochastic computational systems led to the development of appropriate formal models and logical languages. A notable success has been achieved by \emph{probabilistic computational tree logic} \citep[PCTL,][]{HanssonJonsson1994}, and its extensions (e.g., PCTL${}^*$, PRCTL, PCTLK), that are conceived to specify properties of Markov processes. Standard PCTL and PCTL$^{*}$, in particular, are the reference languages for \emph{discrete-time Markov chains} \citep{MC} and \emph{Markov decision processes} \citep{ModelCheckingMDP}, whereas PRCTL and CSL are used, respectively, for \emph{Markov reward models} \citep{andova2003discrete} and \emph{continuous-time Markov chains} \citep{ModelCheckingCTMC}. Other extensions have been also proposed to cope with \emph{hidden Markov models} \citep{ModelCheckingHMM} and \emph{probabilistic interpreted systems} \citep{Primiero2016}.

Despite their wide applicability, PCTL and its extensions suffer from a significant limitation: the probabilities governing the model behaviour require a sharp quantification. This is a serious issue in many real-case scenarios. In computational biology, for instance, non-homogeneous processes are quite common \citep{MCBiologicalSystems2}. Similarly, ignorance about probabilities represents an important challenge for epistemic multi-agent systems. To overcome these limitations, different approaches have been proposed. \emph{Parametric Markov chains} \citep{Parametric2004,ParametricCTMC} model such ignorance by treating transition probabilities as parameters. The applicability of these models is however limited because of the computational complexity of the relative model-checking procedures, which is exponential with respect to the number of states of the model, even for the most advanced procedures based on fraction-free Gaussian elimination \citep{Parametric2020}.

An alternative and less demanding approach is provided by the theory of \emph{imprecise probabilities} \citep{walley1991} and, in particular, by the most recent works on imprecise Markovian processes, which provide a robust approach to the modelling of non-homogeneous Markov processes, as well as
to the modelling of partial ignorance about transition probabilities \citep[e.g.,][]{DeCooman2016}. \cite{Troffaes2013} have outlined the first attempt to extend the framework of probabilistic model checking to imprecise probability models. On the side of properties specification, they replace the standard PCTL probability operator with a new one weighted by an interval whose bounds correspond to the respective lower and upper bounds of probabilistic inferences in an imprecise Markov chain. On the semantic side, they prove that relevant model-checking tasks concerning probabilistic formulae can be reduced to well-known marginalization tasks in imprecise Markov chains. Finally, they provide specific algorithms that exploit marginalization to check whether a model satisfy a given imprecise probabilistic formula.

The contribution of \citet{Troffaes2013} is focused on discrete-time Markov chains. A very natural way to extend such seminal work is to consider an imprecise version of Markov reward models. The present work explores this extension. Imprecise Markov reward models are simply intended as imprecise Markov chains provided with a labelling function that assigns a numerical reward to each state in the model.
Two new robust inference tasks are considered, these being the computation of the bounds of \emph{expected cumulative reward} and of the \emph{reward-bounded} probabilities. The IPCTL language proposed by \citet{Troffaes2013} is therefore extended here including new operators to represent those inferences. We call the language we obtain IPRCTL. We provide the new language with a proper semantics and define satisfiability conditions for the new operators. Hence, we present specific algorithms to compute the relevant inferences specified by those operators. These algorithms are derived by the schema introduced by \citet{TJoensKBC19} to compute robust inferences in imprecise Markov chains. Finally, we outline some considerations about the computational complexity of those new algorithms. Notably shifting from precise to imprecise probabilities does not affect the overall computational complexity of the relevant model-checking procedures.

The paper is organized as follows. In Section \ref{sec:background} we review some basic material. The definitions and algorithms for, respectively, precise and imprecise Markov chains are discussed in Sections \ref{sec:mc} and \ref{sec:imc}. Section \ref{sec:pctl} contains the syntax and semantics of both PCTL and PRCTL, while the imprecise-probabilistic extensions are discussed in Section \ref{sec:ipctl}. Before the conclusions in Section \ref{sec:conc}, we validate these algorithms in Section \ref{sec:case_study} with a case study about the spend-down costs of geriatric medicine departments based on a sensitivity analysis of the results in \citet{MRMGeriatric}.

\section{Background}\label{sec:background}
We first review the necessary background material and notation about the theory of imprecise probability. Let $S$ be a variable taking its values from a finite non-empty set of states $\mathcal{S}$ whose generic elements are denoted by $s\in\mathcal{S}$. A \emph{probability mass function} (PMF, for short) over $S$, denoted by $P(S)$, is defined as a non-negative normalized real map over $\mathcal{S}$. Given a real-valued function $f$ of $S$, i.e., $f:\mathcal{S}\to\mathbb{R}$, its \emph{expectation} based on $P$ is defined as $E[f]:=\sum_{s\in\mathcal{S}} f(s) P(s)$. Notation $P(S'|s):=\{P(s'|s)\}_{s\in\mathcal{S}}$ and $P(S'|S):=\{ P(S'|s)\}_{s\in\mathcal{S}}$ is used instead for conditional probabilities.

A \emph{credal set} (CS) over $S$, denoted as $K(S)$, is a collection of PMFs over $S$. We consider here only finitely generated CSs, i.e., CSs whose convex hull has only a finite number of extreme points. Given a real-valued function $f$ of $\mathcal{S}$, the upper expectation of $f$ with respect to $K(S)$ is defined as $\overline{E}[f]:=\sup_{P(S)\in K(S)} E_P[f]$. The lower expectation $\underline{E}$ is similarly defined. Suprema (infima) of upper (lower) expectations can be equivalently reduced to maxima (mimima) over the extreme points of the CS convexification. Consequently, we can identify a CS with the extreme points of its convex hull. Conditional expectations can be similarly considered.

\section{Markovian Models}
Les us first discuss how \emph{discrete-time Markov chains} can model the behaviour of stochastic, time-homogeneous and memory-less, agents with a finite number of possible states. 

\subsection{(Precise) Markov Chains}\label{sec:mc}
Consider an agent defined over a finite non-empty set of possible states $\mathcal{S}$. The agent evolves  from a state $s\in \mathcal{S}$ to another state $s'\in \mathcal{S}$. Any possible temporal evolution of the agent across time is described by a countable sequence of states that is called a \emph{path} and denoted by $\pi$. Similarly, we use $\Pi$ to denote the set of all possible paths, whereas we use $\pi(t)$ to denote the generic state of the path $\pi$ at time $t\in \mathbb{N}$. The agent is stochastic, meaning that there is a certain degree of \emph{uncertainty} about which path describes its actual evolution. This uncertainty can be measured. To do so, we endow $\Pi$ with a $\sigma$-algebra $\sigma(\Pi)$ and augment it to a probability space $(\Pi,\sigma(\Pi),P)$.
\footnote{In particular, $\sigma(\Pi)$ is the $\sigma$-algebra generated by the cylinder sets, also called \emph{cylinder $\sigma$-algebra} \citep{revuz2008markov}. This allows all the functions we introduce to be measurable.}
Over this probability space, we define a family $\{S_t\}_{t\in \mathbb{N}}$ of categorical stochastic variables $S_t$ such that, for all $t\in \mathbb{N}$, $S_t: \pi\mapsto \pi(t)$. For each $t\in \mathbb{N}$, $P(S_t)$ denotes a PMF that assigns to each $s\in \mathcal{S}$ the probability of $s$ to be the state of the agent at time $t\in \mathbb{N}$. Similarly, for each $t\in \mathbb{N}$, $P(S_{t+1}=s'|S_{t}=s)$ are the conditional probabilities modelling, for each pair of states $s,s'\in \mathcal{S}^2$, the probability of the agent to reach $s'$ at time $t+1$ given that it is in state $s$ at time $t$. Because the agent is memory-less, it satisfies the Markov property, i.e.,
\begin{equation}\label{ref:Markovproperty}
    P(S_{t+1}|S_{t},\ldots,S_{0}) = P(S_{t+1}|S_{t}).
\end{equation}
Furthermore, we assume the behaviour of the agent to be \emph{time-homogeneous}, i.e., $P(S_{t+1}|S_t)$ is the same for all $t\in \mathbb{N}$. Given the Markov property and the time-homogeneity, a compact specification of the agent behaviour is possible in terms of an initial PMF $P(S_0)$ and a transition matrix $T: \mathcal{S}^{2}\mapsto [0,1]$, whose elements are the values in $P(S_{t+1}|S_{t})$ and where the choice of $t$ is arbitrary because of time-homogeneity. Such a model is called here a \emph{precise Markov chain} (PMC) and denoted by $M$.

Typical inferential tasks in PMCs can be computed by means of a \emph{transition operator} $\hat{T}$ mapping a real-valued function $f$ of $\mathcal{S}$ to its (left) scalar product by $T$, i.e.,
\begin{equation}\label{eq:t_primal}
(\hat{T}f)(s):=\sum_{s'\in\mathcal{S}} T(s',s) \cdot f(s') \,,
\end{equation}
for each $s\in\mathcal{S}$. The \emph{dual} $\hat{T}^\dagger$ of this linear operator is obtained by a right scalar product as follows:
\begin{equation}\label{eq:t_dual}
(\hat{T}^\dagger f)(s)  := \sum_{s'\in\mathcal{S}} T(s,s')  \cdot f(s')\,, \end{equation}
for each $s\in \mathcal{S}$. By total probability it is easy to check that $\hat{T}P(S_t)=P(S_{t+1})$ and hence $\hat{T}^{t} P(S_0)=P(S_{t})$. By the notion of conditional expectation, $\hat{T}^\dagger f(S_t)=E_P[f(S_{t+1})|S_t]$, and hence $((\hat{T}^\dagger)^t f(S_0))(s)=E_P[f(S_{t})|S_0=s]$.

We similarly compute the \emph{hitting probability} vector $h_\mathcal{A}^{\leq t}$ for a finite time-horizon $t\in \mathbb{N}$. For each $s\in \mathcal{S}$, and $\mathcal{A}\subseteq \mathcal{S}$, $h_\mathcal{A}^{\leq t}(s)$ is defined as the probability of having at least a state $s_\tau \in \mathcal{A}$ for some $\tau \leq t$, provided that $S_0=s$. Clearly, $h_\mathcal{A}^{\leq 0}=\mathbb{I}_{\mathcal{A}}$, i.e., the indicator function of $\mathcal{A}$ gives the hitting vector for $t=0$ being one for states consistent with $\mathcal{A}$ and zero otherwise. We say that a state $s\in\mathcal{S}$ is \emph{absorbing} if $T(s,s')=0$ for each $s'\neq s$, i.e., once the model is in an absorbing state, the transition probabilities to other states are all zero. Let $T_\mathcal{A}$ denote the transition matrix obtained from $T$ by making absorbing all states $s\in \mathcal{A}$. We can obtain the hitting probability as:
\begin{equation}\label{eq:hitting_primal}
h_\mathcal{A}^{\leq t}(s) = \sum_{s'\in \mathcal{A}} T_\mathcal{A}^t(s,s') = [(\hat{T}^{\dagger})^t \mathbb{I}_{\mathcal{A}}](s)\,,
\end{equation}
for $s \in \mathcal{A}^c:=\mathcal{S}\setminus \mathcal{A}$, while, trivially, $h_\mathcal{A}^{\leq t}(s)=1$ for $s\in\mathcal{A}$. 
The dual of the above computation corresponds to the recursion:
\begin{equation}\label{eq:hitting_dual}
h_{\mathcal{A}}^{\leq t} = \mathbb{I}_{\mathcal{A}} + \mathbb{I}_{\mathcal{A}^c} \hat{T}^\dagger h_{\mathcal{A}}^{\leq t-1}\,,
\end{equation}
for each $t>0$, where sums and products by indicator functions are intended as element-wise operations on arrays.

The unbounded hitting probability vector $h_{\mathcal{A}}$ whose elements are the values of the probability of the agent to reach at least a state $s\in \mathcal{A}$ \emph{evetually in the future} computed for each $s\in \mathcal{S}$, can be achieved as $\lim_{t\to \infty}h^{\leq t}_{\mathcal{A}}$. Proves of the existence of this limit are available in literature, see \citet{revuz2008markov}. Here, simply note that $\lim_{t\to \infty}h^{\leq t}_{\mathcal{A}}$ corresponds to the fixed point of Equation \eqref{eq:hitting_dual}. We remind the reader to classical references, e.g., \citet{revuz2008markov}, for the formal proofs of these results.

Finally, following \citet{katoen2005markov}, we define a \emph{Markov reward model} (MRM) as a PMC paired with a reward function $rew: \mathcal{S} \mapsto \mathbb{N}$. We call the natural number $rew(s)$, the reward of state $s\in\mathcal{S}$, while  $rew$ also denotes the array of all the rewards for the different values of $\mathcal{S}$. The \emph{cumulative} reward of the $t$-th state of a path $\pi$ in the time range $[0,t]$ is intended as:
\begin{equation}\label{eq:cumulative}
Rew(\pi, t):= \sum_{\tau=0}^{t}rew(\pi(\tau))\,.
\end{equation}

\subsection{Imprecise Markov Chains}\label{sec:imc}
The generalization of PMCs to imprecise probability can be achieved in different ways, possibly leading to different inferential results on specific tasks  \citep{krak2019hitting}. The approach we adopt here, also called  \emph{model-theoretic}, is based on a direct sensitivity analysis interpretation  approach): an IMC is intended as a family of PMCs all compatible with the assessments about the system uncertain behaviour. 
Under this interpretation,  
a generalization of a PMC to an IMC can be achieved by replacing $P(S_0)$ with a CS $K(S_0)$ and the transition matrix $T$ with a \emph{credal transition matrix} $\mathcal{T}$ made of conditional CSs, i.e., $\mathcal{T}:=\{K(S'|s)\}_{s\in\mathcal{S}}$ and characterizing the transitions from $S$ to $S'$. Time-homogeneity consists instead in assuming the specification of the (collections of) CSs $K(S_{t+1}|S_t)$ independent of $t$. 
The dual (linear) transition operator in Equation \eqref{eq:t_dual} admits the following (non-linear) extension to IMCs:
\begin{equation}\label{eq:impreciseT}
(\overline{\mathcal{T}}f)(s):= \sup_{T(s,S') \in K(S'|s)} \sum_{s'\in \mathcal{S}} T(s,s') \cdot f(s')\,,
\end{equation}
to be considered for each $s\in\mathcal{S}$, with $K(S'|s)\in\mathcal{T}$ \citep{Troffaes2013}. An analogous \emph{lower} operator $\underline{\mathcal{T}}$ can be defined by replacing the supremum in Equation \eqref{eq:impreciseT} with an infimum. Note that the optimization in Equation \eqref{eq:impreciseT} is a linear programming task whose feasible region is the convex hull of $K(S'|s)$, that in our assumptions can be described by a finite number of linear constraints. It is easy to check that:
\begin{equation}
(\overline{\mathcal{T}}^t f)(s)= \overline{E}(f(S_t)|S_0=s)\,,
\end{equation}
for each $s\in\mathcal{S}$. A similar relation holds for the lower operator and the lower expectation.
As recently shown by \citet{TJoensKBC19}, the recursion in Equation \eqref{eq:hitting_dual} to compute the hitting probabilities in a PMC can be easily generalized to IMCs as follows:
\begin{equation}\label{eq:hitting_dual_imprecise}
\overline{h}_\mathcal{A}^{\leq t}= \mathbb{I}_\mathcal{A} + \mathbb{I}_{\mathcal{A}^c} \overline{\mathcal{T}}\, \overline{h}_{\mathcal{A}}^{\leq t-1}\,,
\end{equation}
with starting point $\overline{h}_{\mathcal{A}}^{\leq 0}=\mathbb{I}_\mathcal{A}$. The upper hitting probability vector $\overline{h}_{\mathcal{A}}^{\leq t}$ is  intended as the upper bound, computed with respect to the joint CS induced by the IMC, of the hitting probability vector defined in Section \ref{sec:imc}. A similar recursion, involving the lower operator and giving the lower hitting probability also holds. As shown by \citet{Troffaes2013}, those recursions can equivalently provide a generalization of Equation \eqref{eq:hitting_primal}:
\begin{equation}\label{eq:hitting_primal_imprecise}
\overline{h}_\mathcal{A}^{\leq t}(s)= 
(\overline{\mathcal{T}}^{t}_\mathcal{A} \mathbb{I}_{\mathcal{A}})(s)\,,
\end{equation}
for each $s\in\mathcal{A}^c$ being instead one for $s\in\mathcal{A}$, where $\overline{\mathcal{T}}_\mathcal{A}$ is such that $\overline{\mathcal{T}}_\mathcal{A}f(s)=f(s)$ for $s\in\mathcal{A}$ and $\overline{T}_\mathcal{A}f(s)=\overline{T}f(s)$ for $s\in\mathcal{A}^c$. 
When paired with a reward function $rew$, the IMC is called \emph{imprecise} MRM (IMRM). The problem of computing lower and upper expectations of the cumulative reward of a path as in Equation \eqref{eq:cumulative} is one of the algorithmic challenges we want to address in the rest of the paper.

\section{Probabilistic Computational Tree Logic}\label{sec:pctl}
We open the discussion by reviewing the syntax (Section \ref{sec:pctlsyntax}) and semantics (Section \ref{sec:pctlsemantics}) of PCTL, the reference language for MC based on PMCs. A demonstrative model-checking task involving PCTL is in Section \ref{sec:toy}. Afterwards, we show how the PCTL syntax has been extended to take into account reward functions in PRCTL (Section \ref{sec:prctlsyntax}), whose semantics is reduced to MRM queries (Section \ref{sec:prctlsemantics}).

\subsection{PCTL Syntax}\label{sec:pctlsyntax}
The PCTL syntax is recursively defined as follows:
\begin{align}
\phi:= & \top \mid p \mid \neg \phi \mid \phi_1 \land \phi_2 \mid P_{\nabla{b}}\psi\,,  \label{eq:pctlsyntax1}\\
\psi:= & \bigcirc\phi \mid \phi_1\bigcup\phi_2 \mid \phi_1\bigcup^{\leq t}\phi_2
\,.\label{eq:pctlsyntax2}
\end{align}
The language includes the standard notation $\top$ for \emph{true}, atoms (such as $p$) and standard Boolean formulae, whose meaning is the same as in standard propositional logic. It also includes path formulae denoted by $\psi$ and representing properties of paths with the following informal reading \citep{MC}:
\begin{itemize}
\item $\bigcirc\phi$ means that \emph{in the next state of the path $\phi$ hold};
\item $\phi_1\bigcup^{\leq t}\phi_2$ means that \emph{$\phi_2$ holds at a certain  time $\tau\leq t$ and $\phi_1$ holds in all the previous states of the path};
\item $\phi_1\bigcup\phi_2$ means that \emph{$\phi_2$ holds eventually along the path and $\phi_1$ holds in all the previous states of the path}.
\end{itemize}
Finally, for the probability formulae, where $b\in [0,1]$ and $\nabla{}\in\{<,\leq,=,\geq,>\}$, $P_{\nabla{b}}\psi$ means that \emph{there is a probability $\nabla{b}$ to reach a path satisfying $\psi$}.

\subsection{PCTL Semantics}\label{sec:pctlsemantics}
To present PCTL semantics we first augment standard PMCs (Section~\ref{sec:mc}) with a finite non-empty set of atomic propositions $AP$ and a \emph{labelling} function $l: \mathcal{S} \mapsto 2^{AP}$ that assigns to each state $s\in \mathcal{S}$ a set of propositions $l(s)\subseteq AP$. The resulting model is called \emph{labelled PMC}. In the following, by PMC we always denote a labelled PMC.

\paragraph{Boolean Formulae.} Given PMC $M$ and state $s\in \mathcal{S}$, the semantics for Boolean formulae is as follows:
\begin{align}
M,s & \models \top\,\,\forall s\in \mathcal{S}\,;\\
M,s & \models p \,\,\mathrm{iff}\,\, p\in l(s)\,;\\
M,s & \models \phi_1\land \phi_2 \,\,\mathrm{iff}\,\, M,s\models \phi_1 \,\,\mathrm{and}\,\, M,s  \models \phi_2\,;\\
M,s & \models \neg\phi \,\,\mathrm{iff}\,\, M,s \not\models \phi\,.
\end{align}

\paragraph{Path Formulae.} Given PMC $M$ and path $\pi$, the following conditions hold:
\begin{align}
 M,\pi & \models \bigcirc\phi \,\,\mathrm{iff}\,\, M,\pi(1)\models \phi\,;\\
 M,\pi & \models \phi_1\bigcup^{\leq t}\phi_2 \,\,\mathrm{iff}\,\, \exists \tau\leq t:
\begin{array}{l}
M,\pi(\tau)\models\phi_2\,,\\ M,\pi(\tau')\models \phi_1\,,\\ \forall\tau': 0\leq \tau'<\tau\,;
\end{array}\\
M,\pi & \models \phi_1\bigcup\phi_2 \,\,\mathrm{iff}\,\, \exists \tau\geq 0:
\begin{array}{l}
M,\pi(\tau)\models \phi_2\,,\\
M,\pi(\tau')\models \phi_1\,,\\
\forall \tau': 0\leq \tau'<\tau\,.
\end{array}
\end{align}

The \emph{check} of the models with respect to Boolean formulae can be achieved by SAT solvers \citep{davis1960computing}, while the \emph{parsing-tree} technique is used instead for path formulae \citep{MC}. 

\paragraph{Probability Formulae.} Given PMC $M$ and state $s\in \mathcal{S}$ the following condition holds:
\begin{equation}\label{eq:probformula}
M,s\models P_{\nabla{b}}\psi \,\,\mathrm{iff}\,\,
P(s\models \psi)\nabla{b}\,,
\end{equation}
provided that $P(s\models \psi)$ is the probability of the PMC to reach a path $\pi\in \Pi$ such that $\pi\models \psi$ given that $S_0 = s$ \citep{MC}.
For each one of the possible values of $\psi$, the computation of $P(s\models \psi)$ can be reduced to a PMC inference as in the following. Let us begin by considering the case $\psi:=\bigcirc\phi$. The computation of $P(s\models \psi)$ can be achieved simply defining $\Phi:=\{ s'\in\mathcal{S}: M,s' \vDash \phi \}$ and computing the trivial inference:
\begin{equation}\label{eq:conditional}
P(S_{1}\in \Phi|S_{0} = s):=(\hat{T}^\dagger\mathbb{I}_{\Phi})(s) = \sum_{s'\in \Phi}T(s,s')\,.
\end{equation}

We now consider the case $\psi := \phi_1\bigcup^{\leq t}\phi_2$. To compute $P(s\models \phi_1\bigcup^{\leq t}\phi_2)$, we first define $\Phi_1$ and $\Phi_2$ as the subsets of $\mathcal{S}$ satisfying, respectively, $\phi_1$ and $\phi_2$. The probability in Equation \eqref{eq:probformula}, hence, can be computed as the hitting probability of event $\Phi_2|\Phi_1$, that is the set of all $s\in \mathcal{S}$ such that $s\in \Phi_2$ and all the states $s'\in \mathcal{S}$ visited before reaching $s$ are in $\Phi_1$. We denote such hitting probability as $h_{\Phi_2|\Phi_1}^{\leq t}$. A recursion analogous to that in Equation \eqref{eq:hitting_dual} is obtained by multiplying the complement of the hitting event $\Phi_2$ by the indicator of $\Phi_1$, thus obtaining the indicator of the set difference , i.e.,
\begin{equation}\label{eq:cond_hitting}
h_{\Phi_2|\Phi_1}^{\leq \tau}:= \mathbb{I}_{\Phi_2} + \mathbb{I}_{\Phi_1\setminus \Phi_2}\hat{T}^{\dagger} h_{\Phi_2|\Phi_1}^{\leq \tau-1}\,,
\end{equation}
for each $\tau=1,\ldots,t$, with $h_{\Phi_2|\Phi_1}^{\leq 0}:=\mathbb{I}_{\Phi_2}$. 
Case $\psi := \phi_1\bigcup \phi_2$ is analogous to $\phi_1\bigcup^{\leq t}\phi_2$ when $t\to \infty$. The value of $P(s\models \phi_1\bigcup\phi_2)$, indeed, corresponds to $\lim_{t\to \infty}h_{\Phi_2|\Phi_1}^{\leq t}$. Remind that, this limit exists and it corresponds to the fixed point of the schema in Equation \eqref{eq:cond_hitting} \citep{revuz2008markov}. For this reason, an easy way to approximate $\lim_{t\to \infty}h_{\Phi_2|\Phi_1}^{\leq t}$ consists in iterating the schema in Equation \eqref{eq:cond_hitting} until convergence. Notice that, limited to PMCs, there exist other procedures that allow to directly compute the value of $P(s\models \phi_1\bigcup\phi_2)$ solving a system of linear equations through linear programming \citep[pp. 761-762]{MC}. 

\subsection{A PCTL Example}\label{sec:toy}

\begin{example}[\citet{MC}]\label{ex:toy}
Consider a simple communication model operating with a single channel. The channel is error-prone, meaning that messages can be lost. In this particular example, the (four) states of such model $M$ are in one-to-one correspondence with the atomic propositions, i.e.,  $\mathcal{S}=AP:= \{\mathrm{start}, \mathrm{try}, \mathrm{lost}, \mathrm{delivered}\}$. Transition probabilities are shown in Figure~\ref{fig:4states} as label arrows, impossible transitions correspond to missing edges.

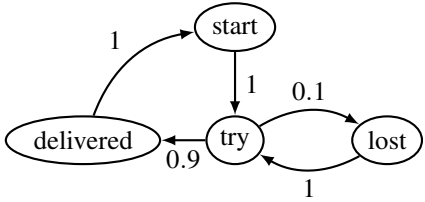
\begin{figure}[htp!]\label{fig:4states}
\centering
\begin{tikzpicture}[style=thick,shorten >=0.1pt,->,every 
path/.style={-latex}]
\tikzstyle{vertex}=[ellipse,draw,fill=white,minimum size=18pt,inner sep=2pt]

\node[vertex]  (1)  at  (3, 2.5) {start};  
\node[vertex]  (2)  at  (1, 1) {delivered};
\node[vertex]  (3)  at  (3,1) {try}; 
\node[vertex]  (4)  at  (5, 1) {lost};  
\path[->](1)  edge [] node[right]  {$1$}  (3);
\path[->](3)  edge [] node[below]  {$0.9$}  (2);
\path[->](2)  edge [bend left] node[above left]  {$1$}  (1);
\path[->](3)  edge [bend left] node[above]  {$0.1$}  (4);
\path[->](4)  edge [bend left] node[below]  {$1$}  (3);
\end{tikzpicture}
\caption{Transitions in a four-state PMC.}
\end{figure}

$M$ is compliant if and only if ``\emph{the probability of a message to be lost within seven time steps is smaller than or equal to $0.25$}''. As the starting state for $M$ is by construction $s=\mathrm{start}$, this corresponds to the following PCTL formula:
\begin{equation}
M,\mathrm{\{start\}} \models P_{\leq 0.25}\top\bigcup^{\leq 7}(\{\mathrm{lost}\})\,,
\end{equation}
whose semantics is such as in Equation \eqref{eq:probformula}.
Yet, the task reduces to compute $h^{\leq 7}_{\{\mathrm{lost}\}}(\mathcal{S}|\{\mathrm{start}\})$.
Its computation can be achieved either by the recursion in Equation~\eqref{eq:hitting_dual} or by the matrix product in Equation \eqref{eq:hitting_primal}, both leading to the numerical values in Table \ref{tab:hitting}. Since the corresponding values is $h^{\leq 7}_{\{\mathrm{lost}\}}(\mathrm{start})=0.190 \leq 0.25$, the system satisfies the requirement and can be considered compliant.

\begin{table}[htp!]
\centering
{\small\begin{tabular}{@{}lcccc@{}}
\toprule
&\multicolumn{4}{c}{$s$}\\
$t$&start&deliv.&try&lost\\
\midrule
0& 0.000& 0.000 &0.000&1.000\\
1& 0.000& 0.000 &0.100&1.000\\
2& 0.100& 0.000 &0.100&1.000\\
3& 0.100& 0.100 &0.100&1.000\\
4& 0.100& 0.100 &0.190&1.000\\
5& 0.190& 0.100 &0.190&1.000\\
6& 0.190& 0.190 &0.190&1.000\\
7& {\bf 0.190}& 0.190 &0.271&1.000\\
\bottomrule
\end{tabular}}
\caption{Hitting probabilities $h_{\{lost\}}^{\leq t}(s)$ for $t\leq 7$.}\label{tab:hitting}
\end{table}
\end{example}

\subsection{PRCTL Syntax}\label{sec:prctlsyntax}
PRCTL \citep{katoen2005markov} is a PCTL extension achieved by augmenting Equations \eqref{eq:pctlsyntax1} and \eqref{eq:pctlsyntax2} as follows:
\begin{align}\label{eq:prctl_syntax1}
\phi:= & \dots \mid E_{\nabla{r}}\phi\,,\\
\label{eq:prctl_syntax2}
\psi:= & \dots \mid \phi_1\bigcup_{\leq r}\phi_2\,,
\end{align}
where $\nabla{}:=\{<,\leq,=,\geq,>\}$ and $r \in \mathbb{N}$.
\footnote{Following \citet{MC}, we assume rewards to be natural instead of real numbers. This assumption is very common in the field of model checking because it aids to avoid possible issues concerning the overall computational complexity of the main tasks.} The informal reading of the state formula $E_{\nabla{r}}\phi$ is that \emph{the expected cumulative reward earned until reaching a state that satisfies $\phi$ is $\nabla{r}$}. The \emph{bounded reward} path formula $\phi_1\bigcup_{\leq r}\phi_2$ means instead that \emph{there exists $t\geq 0$ such that $\pi(t)$ satisfies $\phi_2$ while all the previous states along the path satisfy $\phi_1$ and the cumulative reward of path $\pi$ from $0$ to $t$ is less than or equal to $r$}.

\subsection{PRCTL Semantics}\label{sec:prctlsemantics}
To define the PRCTL semantics we add three new satisfiability conditions to those in Section \ref{sec:pctlsemantics}.
\paragraph{Bounded Reward.} Given MRM $(M, rew)$ and path $\pi$, the following condition holds:
\begin{equation}
    (M,rew),\pi\models \phi_1\bigcup_{\leq r}\phi_2 \,\, \mathrm{iff}\,\,  \exists \tau: 
    \begin{array}{l}
    (M,rew),\pi(\tau)\vDash \phi_2\,, \\
    (M,rew),\pi(\tau')\vDash \phi_1, \forall \tau'<\tau \,,\\
    Rew(\pi, \tau)\leq r\,.
    \end{array}
\end{equation}
The procedure for this formula is analogous to that of the PCTL path formulae. 
\paragraph{Expected Cumulative Reward.} Given MRM $(M, rew)$ and state $s\in \mathcal{S}$, the following condition holds:
\begin{equation}\label{eq:exprewformula}
(M,rew),s\models E_{\nabla{r}}\phi \,\,\mathrm{iff}\,\,E[Rew^{\leq t}_\Phi](s) \nabla{r}\,.
\end{equation}
Notice that, when $t\to \infty$, i.e., when no state satisfying $\phi$ is reached until a finite time $t\in \mathbb{N}$, by definition $E[Rew^{\leq t}_\Phi](s) = +\infty$. 
Instead, for each $t\in \mathbb{N}$ and $s\in\mathcal{S}$ the expected cumulative reward is defined as follows:
\begin{equation}\label{eq:expcumrew}
E[Rew_\Phi^{\leq t}](s) := \sum_{\substack{\pi \in \mathrm{Paths}(s):\\\exists\tau\leq t: \pi(\tau) \models \phi}} \!\!\!\!\!\!Rew(\pi, \tau) \cdot P(\pi)\,,
\end{equation}
That is, an expected value for paths starting in $S_0=s$ and reaching $\Phi$ at some time step $\tau \leq t$. A recursion analogous to that in Equation \eqref{eq:hitting_dual} for the hitting event $\Phi$ can be therefore derived for the quantity in Equation \eqref{eq:expcumrew}. The only difference is that, if event $\Phi$ is achieved for $S_0=s$, the value one in the indicator function should be multiplied by the corresponding reward for $s$. If this is not the case, the expected rewards of the previous time step should be propagated by the transition operator, and increased by the original rewards, i.e.,
\begin{equation}
E[Rew_\Phi^{\leq\tau}] = \mathbb{I}_{\Phi}\cdot rew + \mathbb{I}_{\Phi^c} ( rew + \hat{T}^{\dagger} E[Rew^{\leq \tau-1}_{\Phi}])\,,
\end{equation}
and hence:
\begin{equation}\label{eq:recursion_expcumrew}
E[Rew_\Phi^{\leq\tau}] = rew + \mathbb{I}_{\Phi^c} \hat{T}^{\dagger} E[Rew^{\leq \tau-1}_{\Phi}]\,,
\end{equation}
where both equations are array relations to be considered for each $\tau=1,\ldots,t$, sums and products should be intended as element-wise operations on arrays, and $rew$ is the array of (state) rewards defined in Section \ref{sec:mc}. 
Note that, when $rew=\mathbb{I}_{\Phi}$, Equation \eqref{eq:recursion_expcumrew} becomes Equation \eqref{eq:hitting_dual}, i.e., if rewards are one on the hitting event and zero otherwise, the expected cumulative reward equals the hitting probability.

\paragraph{Bounded-Reward Probability.}
Given MRM $(M,rew)$ and state $s \in \mathcal{S}$, the following condition holds:
\begin{equation}\label{eq:bounded_rew_semantics}
(M,rew),s\models P_{\nabla{b}}\phi_1\bigcup_{\leq r}\phi_2 \,\,\mathrm{iff}\,\, P(s\models \phi_1\bigcup_{\leq r}\phi_2) \nabla{b}\,,
\end{equation}
where, as in Equation \eqref{eq:probformula}, $P(s\models \phi_1\bigcup_{\leq r}\phi_2)$ is the probability of the PMC to reach a path $\pi\models \phi_1\bigcup_{\leq r}\phi_2$ given that $S_0 = s$.
This can be intended as a hitting probability for event $\Phi_2$ with respect to state $s\in \mathcal{S}$ provided that all the states visited before reaching $\Phi_2$ satisfy $\phi_1$, and the expected cumulative reward earned before reaching $\Phi_2$ is less than or equal to $r$. This remark allows to derive a recursion analogous to that in Equation \eqref{eq:cond_hitting}, that also takes into account the reward earned after any iteration. For each $\tau\geq 0$, $\rho \leq r$ and $s\in \mathcal{S}$, we denote by $x^{\leq \tau,\rho}_{\Phi_2|\Phi_1}(s)$ the hitting probability of reaching $\Phi_2$ for some $\tau\geq 0$ provided that all the states visited before reaching $\Phi_2$ are in $\Phi_1$ and the expected cumulative reward earned until reaching $\Phi_2$ is $\leq r$. Trivially, for $\tau=0$:
\begin{equation}\label{eq:init_bounded_prob2}
    x^{\leq 0,\rho}_{\Phi_2|\Phi_1}(s):=
    \begin{cases}
    1\,\, \mathrm{if}\,\, s\in \Phi_2\,\, \mathrm{and}\,\, rew(s)\leq \rho\,,\\
    0\,\, \mathrm{otherwise}.
    \end{cases}
\end{equation}
In array notation, this rewrites as follows:
\begin{equation}\label{eq:init_bounded_prob}
x^{\leq 0, \rho}_{\Phi_2|\Phi_1} :=
\mathbb{I}_{\mathcal{S}_{\rho}^{rew}}\cdot
\mathbb{I}_{\Phi_2}\,,
\end{equation}
where $\mathcal{S}_{\rho}^{rew}:=\{s\in \mathcal{S}: rew(s)\leq \rho\}$. 
For $\tau>0$, the following recursion holds \citep{MC}:
\begin{equation}\label{eq:Bounded}
x^{\leq \tau,\rho}_{\Phi_2|\Phi_1}(s):=
\begin{cases}
1 \,\,\mathrm{if}\,\, s\in \Phi_2\,\, \mathrm{and}\,\, rew(s)\leq \rho\,,\\
0 \,\,\mathrm{if}\,\, rew(s)> \rho\,,\\
\sum_{s'\in \Phi_1}T(s,s') x^{\leq \tau-1,\rho- rew(s')}_{\Phi_2|\Phi_1}(s')\,\, \mathrm{otherwise}\,,
\end{cases}
\end{equation}
i.e.,
\begin{equation}\label{ref:RecursionBoundedReward}
x^{\leq \tau, \rho}_{\Phi_2|\Phi_1}:= 
\mathbb{I}_{\mathcal{S}_\rho^{rew}} \left(
\mathbb{I}_{\Phi_2}+ 
\mathbb{I}_{\Phi_1\setminus\Phi_2}\cdot \hat{T}^{\dagger}\chi^{\leq\tau-1,\rho} 
\right)\,,
\end{equation}
where, for each $s\in \mathcal{S}$:
\begin{equation}\label{eq:chi}
\chi^{\leq\tau-1,\rho}(s):= 
\left\{ 
\begin{array}{ll}
1& \mathrm{if} \,\rho\leq rew(s)\,,\\
x^{\leq \tau-1, \rho-rew(s)}_{\Phi_2|\Phi_2}(s)&\mathrm{otherwise} 
\,.\end{array}\right.
\end{equation}
This defines a simple algorithmic scheme where  
$x^{\leq \tau, \rho}_{\Phi_2|\Phi_2}(s)$ should be computed as in Equation \eqref{eq:Bounded} for each $s\in \mathcal{S}$ and $\rho \leq r$ before moving to the subsequent value of $\tau$. According to Equation \eqref{eq:Bounded}, the recursion is blocked for each $s\in \Phi_2$, with the final probability being equal to one, and for each $s\in \mathcal{S}: rew(s)>\rho$, with the final probability being equal to zero. The recursion is eventually blocked when, for all the reached states $s\in \mathcal{S}$ it holds that either $s\in \Phi_2$ or $rew(s) > \rho$. This always happens within a finite time horizon $\tau \in \mathbb{N}$ because, at each further recursive step, the reward threshold $\rho$ is reduced, for each $s'\in \mathcal{S}$, of a value $rew(s')$ as specified by Equation \eqref{eq:chi}. The final value of $P(s\models \phi_1\bigcup_{\leq r}\phi_2)$ is then equal to $x^{\leq \tau,r}_{\Phi_2|\Phi_1}$ where $\tau\in \mathbb{N}$ is the time step at which \eqref{ref:RecursionBoundedReward} is eventually blocked.

\section{Towards an Imprecise PRCTL}\label{sec:ipctl}
In this section we discuss how both PCTL and PRCTL can be extended to an imprecise-probabilistic setting. An \emph{imprecise} PCTL (IPCTL) syntax can be obtained by replacing PCTL probability operator $P_{\nabla{b}}$ with its lower (upper) variant   $\underline{P}_{\nabla{b}}$ ($\overline{P}_{\nabla{b}}$) (Section \ref{sec:ipctlsyntax}). The semantics (Section \ref{sec:ipctlsemantics}) is consequently defined as in Section \ref{sec:pctlsemantics} with the checking tasks performed in the corresponding IMCs by means of the inference algorithms described in Section \ref{sec:imc}.

IPCTL has been proposed by \citet{Troffaes2013}. Here, we take advantage of the alternative approaches to the computation of the lower and upper hitting probabilities recently proposed by \citet{TJoensKBC19} and corresponding to Equation \eqref{eq:hitting_dual_imprecise}. Compared to the absorbing-state approach in Equation \eqref{eq:hitting_primal_imprecise}, proposed by \citet{Troffaes2013}, the more recent approach allows to easily achieve an analogous computation for the lower and upper expected cumulative rewards in the underlying IMRM, thus providing an imprecise-probabilistic version of Equation \eqref{eq:recursion_expcumrew}. This is the key to define an \emph{imprecise} PRCTL (IPRCTL) whose syntax (Section \ref{sec:iprctl_syntax}), semantics and model checking (Section \ref{sec:iprctl_semantics}) represent the main contribution of this work.

\subsection{IPCTL Syntax}\label{sec:ipctlsyntax}
The non-probabilistic syntax of IPCTL coincides with that of PCTL. Accordingly, to define IPCTL syntax, we keep the non-probabilistic specification in Equation \eqref{eq:pctlsyntax2} and, in Equation  \eqref{eq:pctlsyntax1}, replace $P_{\nabla{b}}\phi$ with:
\begin{equation}
\overline{P}_{\nabla{b}}\psi \mid \underline{P}_{\nabla{b}}\psi\,.
\end{equation}
The informal reading of $\overline{P}_{\nabla{b}}\psi$ is  \emph{the upper bound of the probability to reach a path that satisfies $\psi$ is $\nabla{b}$}. A similar reading for the lower bound is associated with $\underline{P}_{\nabla{b}}\psi$. Note that IPCTL coincides with PCTL for $\underline{P} = \overline{P}$.

\subsection{IPCTL Semantics}\label{sec:ipctlsemantics}
The only difference between IPCTL and PCTL semantics is in the probability formulae. Given labelled IMC $\mathcal{M}$ and state $s\in\mathcal{S}$, Equation \eqref{eq:probformula} becomes:
\begin{equation}\label{eq:iprobformula}
\mathcal{M},s\models \overline{P}_{\nabla{b}}\psi \,\,\mathrm{iff}\,\, \overline{P}(s\models \psi) \nabla{b}\,,
\end{equation}
and analogously for $\underline{P}$. As in Section \ref{sec:pctlsemantics} for PCTL, the satisfiability check in Equation \eqref{eq:iprobformula} leads to different inferential tasks in $\mathcal{M}$ depending on $\psi$.

For $\psi:= \bigcirc\phi$, following Equation \eqref{eq:conditional}, the lower and upper bounds of $P(S_{1}\in \Phi|S_{0} = s)$ should be computed. An imprecise-probabilistic version of this equation is achieved by replacing the linear operator in Equation \eqref{eq:t_dual} with its non-linear analogous in Equation \eqref{eq:impreciseT}, i.e.,
\begin{equation}
\overline{P}(S_1\in \Phi|S_0 = s) = 
(\overline{\mathcal{T}} \mathbb{I}_{\Phi})(s)\,,
\end{equation}
and analogously for $\underline{P}$ and $\underline{\mathcal{T}}$. 
The case $\psi:= \phi_1\bigcup^{\leq t}\phi_2$ requires instead the computation of the lower and upper bounds of the conditional hitting probability vector $h^{\leq t}_{\Phi_2|\Phi_1}$. Exactly as Equation \eqref{eq:cond_hitting} was obtained as a conditional version of Equation \eqref{eq:hitting_dual}, from Equation \eqref{eq:hitting_dual_imprecise} we can obtain the recursion:
\begin{equation}\label{eq:cond_lower_hitting}
\overline{h}_{\Phi_2|\Phi_1}^{\leq \tau}:= \mathbb{I}_{\Phi_2} + \mathbb{I}_{\Phi_1\setminus \Phi_2} \overline{\mathcal{T}} \,\overline{h}_{\Phi_2|\Phi_1}^{\leq \tau-1}\,,
\end{equation}
for each $\tau=1,\ldots,t$, and analogously for the upper bound, with the same initialization for both cases, i.e.,
\begin{equation}\label{eq:init_imp_hitting}
\underline{h}_{\Phi_2|\Phi_1}^{\leq 0}=
\overline{h}_{\Phi_2|\Phi_1}^{\leq 0}= \mathbb{I}_{\Phi_2}\,.
\end{equation}
As for PMCs, case $\psi:=\phi_1\bigcup\phi_2$ can be treated analogously to $\phi_1\bigcup^{\leq t}\phi_2$ when $t\to \infty$. The value of $\overline{P}(s\models \phi_1\bigcup\phi_2)$ hence corresponds to $\lim_{t\to \infty}\overline{h}^{\leq t}_{\Phi_2|\Phi_1}$. Proposition 16 in \citep{krak2019hitting} proves that such limit exists and corresponds to the fixed point of $h^{\leq \tau}_{A\subseteq \mathcal{S}}$. Notice that, the result can be considered valid also for the recursion in Equation \eqref{eq:cond_lower_hitting} because the additional condition that all the states visited before reaching $\Phi_2$ are in $\Phi_1$ does not alter the validity of the proof. Notice also that the result by \citet{krak2019hitting} is obtained within a  game-theoretic approach to IMCs, which is different from the sensitivity-analysis approach we adopt here. However, as the author clearly points out, the result is to be considered valid for all the foundational approaches to IMCs \citep{krak2019hitting}.

Regarding computational complexity, the linear programming tasks in Equation \eqref{eq:cond_lower_hitting} take only polynomial time with respect to $|\mathcal{S}|$, while the maximum number of iterations is $t$. This shows that shifting to imprecise probabilities does not affect the overall computational complexity of the task. The only computational issue with computing $\overline{P}(s\models \phi)$ (as well as $\underline{P}(s\models \phi)$) concerns the nesting depth of $\phi$, i.e., the number of iterated nested formulae in $\phi$. The overall complexity of computing $\overline{P}(s\models \phi)$, indeed, is exponential with respect to the nesting depth of $\phi$. However, the same holds in the precise case and this does not significantly affect applications, where small nesting depths are typically considered.

\subsection{A IPCTL Example}\label{sec:itoy}
\begin{figure}[htp!]
\centering
\begin{tikzpicture}[scale=0.6]
\begin{axis}[
tick align=outside,
tick pos=left,
x grid style={white!69.0196078431373!black},
xlabel={\(\displaystyle t\)},
xmin=-0.3, xmax=6.3,
xtick style={color=black},
xticklabels={1,1,2,3,4,5,6,7},
y grid style={white!69.0196078431373!black},
ylabel={\(\displaystyle h_{\{\mathrm{lost}\}}^{\leq t}(\mathrm{start})\)},
ymin=-0.0125, ymax=0.2625,
ytick style={color=black},
ytick={-0.05,0,0.05,0.1,0.15,0.2,0.25,0.3},
yticklabels={−0.05,0.00,0.05,0.10,0.15,0.20,0.25,0.30}
]

\path [draw=black, fill=yellow, opacity=0.5]
(axis cs:0,0)
--(axis cs:0,0)
--(axis cs:1,0.097)
--(axis cs:2,0.097)
--(axis cs:3,0.097)
--(axis cs:4,0.184591)
--(axis cs:5,0.184591)
--(axis cs:6,0.184591)
--(axis cs:6,0.237871)
--(axis cs:6,0.237871)
--(axis cs:5,0.237871)
--(axis cs:4,0.237871)
--(axis cs:3,0.127)
--(axis cs:2,0.127)
--(axis cs:1,0.127)
--(axis cs:0,0)
--cycle;

\path [draw=black, fill=orange, opacity=0.5]
(axis cs:0,0)
--(axis cs:0,0)
--(axis cs:1,0.098)
--(axis cs:2,0.098)
--(axis cs:3,0.098)
--(axis cs:4,0.186396)
--(axis cs:5,0.186396)
--(axis cs:6,0.186396)
--(axis cs:6,0.222076)
--(axis cs:6,0.222076)
--(axis cs:5,0.222076)
--(axis cs:4,0.222076)
--(axis cs:3,0.118)
--(axis cs:2,0.118)
--(axis cs:1,0.118)
--(axis cs:0,0)
--cycle;

\path [draw=black, fill=red, opacity=0.5]
(axis cs:0,0)
--(axis cs:0,0)
--(axis cs:1,0.099)
--(axis cs:2,0.099)
--(axis cs:3,0.099)
--(axis cs:4,0.188199)
--(axis cs:5,0.188199)
--(axis cs:6,0.188199)
--(axis cs:6,0.206119)
--(axis cs:6,0.206119)
--(axis cs:5,0.206119)
--(axis cs:4,0.206119)
--(axis cs:3,0.109)
--(axis cs:2,0.109)
--(axis cs:1,0.109)
--(axis cs:0,0)
--cycle;


\addplot [semithick, black, dashed, forget plot]
table {%
-0.3 0.25
6.3 0.25
};
\end{axis}
\end{tikzpicture}
\begin{tikzpicture}[scale=0.6]
\begin{axis}[
legend cell align={left},
legend style={
  fill opacity=0.8,
  draw opacity=1,
  text opacity=1,
  at={(0.97,0.03)},
  anchor=south east,
  draw=white!80!black
},
tick align=outside,
tick pos=left,
x grid style={white!69.0196078431373!black},
xlabel={\(\displaystyle t\)},
xmin=-5, xmax=105,
xtick style={color=black},
xticklabels={1,,20,,60,,100,},
y grid style={white!69.0196078431373!black},
ylabel={\(\displaystyle h_{\{\mathrm{lost}\}}^{\leq t}(\mathrm{start})\)},
ymin=-0.0495063100455576, ymax=1.03963251095671,
ytick style={color=black}
]
\path [draw=black, fill=yellow, opacity=0.5]
(axis cs:0,0)
--(axis cs:0,0)
--(axis cs:1,0.097)
--(axis cs:2,0.097)
--(axis cs:3,0.097)
--(axis cs:4,0.184591)
--(axis cs:5,0.184591)
--(axis cs:6,0.184591)
--(axis cs:7,0.263685673)
--(axis cs:8,0.263685673)
--(axis cs:9,0.263685673)
--(axis cs:10,0.335108162719)
--(axis cs:11,0.335108162719)
--(axis cs:12,0.335108162719)
--(axis cs:13,0.399602670935257)
--(axis cs:14,0.399602670935257)
--(axis cs:15,0.399602670935257)
--(axis cs:16,0.457841211854537)
--(axis cs:17,0.457841211854537)
--(axis cs:18,0.457841211854537)
--(axis cs:19,0.510430614304647)
--(axis cs:20,0.510430614304647)
--(axis cs:21,0.510430614304647)
--(axis cs:22,0.557918844717096)
--(axis cs:23,0.557918844717096)
--(axis cs:24,0.557918844717096)
--(axis cs:25,0.600800716779538)
--(axis cs:26,0.600800716779538)
--(axis cs:27,0.600800716779538)
--(axis cs:28,0.639523047251923)
--(axis cs:29,0.639523047251923)
--(axis cs:30,0.639523047251923)
--(axis cs:31,0.674489311668486)
--(axis cs:32,0.674489311668486)
--(axis cs:33,0.674489311668486)
--(axis cs:34,0.706063848436643)
--(axis cs:35,0.706063848436643)
--(axis cs:36,0.706063848436643)
--(axis cs:37,0.734575655138289)
--(axis cs:38,0.734575655138289)
--(axis cs:39,0.734575655138289)
--(axis cs:40,0.760321816589875)
--(axis cs:41,0.760321816589875)
--(axis cs:42,0.760321816589875)
--(axis cs:43,0.783570600380657)
--(axis cs:44,0.783570600380657)
--(axis cs:45,0.783570600380657)
--(axis cs:46,0.804564252143733)
--(axis cs:47,0.804564252143733)
--(axis cs:48,0.804564252143733)
--(axis cs:49,0.823521519685791)
--(axis cs:50,0.823521519685791)
--(axis cs:51,0.823521519685791)
--(axis cs:52,0.840639932276269)
--(axis cs:53,0.840639932276269)
--(axis cs:54,0.840639932276269)
--(axis cs:55,0.856097858845471)
--(axis cs:56,0.856097858845471)
--(axis cs:57,0.856097858845471)
--(axis cs:58,0.87005636653746)
--(axis cs:59,0.87005636653746)
--(axis cs:60,0.87005636653746)
--(axis cs:61,0.882660898983327)
--(axis cs:62,0.882660898983327)
--(axis cs:63,0.882660898983327)
--(axis cs:64,0.894042791781944)
--(axis cs:65,0.894042791781944)
--(axis cs:66,0.894042791781944)
--(axis cs:67,0.904320640979095)
--(axis cs:68,0.904320640979095)
--(axis cs:69,0.904320640979095)
--(axis cs:70,0.913601538804123)
--(axis cs:71,0.913601538804123)
--(axis cs:72,0.913601538804123)
--(axis cs:73,0.921982189540123)
--(axis cs:74,0.921982189540123)
--(axis cs:75,0.921982189540123)
--(axis cs:76,0.929549917154731)
--(axis cs:77,0.929549917154731)
--(axis cs:78,0.929549917154731)
--(axis cs:79,0.936383575190722)
--(axis cs:80,0.936383575190722)
--(axis cs:81,0.936383575190722)
--(axis cs:82,0.942554368397222)
--(axis cs:83,0.942554368397222)
--(axis cs:84,0.942554368397222)
--(axis cs:85,0.948126594662692)
--(axis cs:86,0.948126594662692)
--(axis cs:87,0.948126594662692)
--(axis cs:88,0.953158314980411)
--(axis cs:89,0.953158314980411)
--(axis cs:90,0.953158314980411)
--(axis cs:91,0.957701958427311)
--(axis cs:92,0.957701958427311)
--(axis cs:93,0.957701958427311)
--(axis cs:94,0.961804868459862)
--(axis cs:95,0.961804868459862)
--(axis cs:96,0.961804868459862)
--(axis cs:97,0.965509796219255)
--(axis cs:98,0.965509796219255)
--(axis cs:99,0.965509796219255)
--(axis cs:100,0.968855345985987)
--(axis cs:100,0.990126200911153)
--(axis cs:100,0.990126200911153)
--(axis cs:99,0.98868980631289)
--(axis cs:98,0.98868980631289)
--(axis cs:97,0.98868980631289)
--(axis cs:96,0.987044451675704)
--(axis cs:95,0.987044451675704)
--(axis cs:94,0.987044451675704)
--(axis cs:93,0.985159738460142)
--(axis cs:92,0.985159738460142)
--(axis cs:91,0.985159738460142)
--(axis cs:90,0.983000845887906)
--(axis cs:89,0.983000845887906)
--(axis cs:88,0.983000845887906)
--(axis cs:87,0.980527887615013)
--(axis cs:86,0.980527887615013)
--(axis cs:85,0.980527887615013)
--(axis cs:84,0.977695174816739)
--(axis cs:83,0.977695174816739)
--(axis cs:82,0.977695174816739)
--(axis cs:81,0.974450372069575)
--(axis cs:80,0.974450372069575)
--(axis cs:79,0.974450372069575)
--(axis cs:78,0.970733530434794)
--(axis cs:77,0.970733530434794)
--(axis cs:76,0.970733530434794)
--(axis cs:75,0.966475979879489)
--(axis cs:74,0.966475979879489)
--(axis cs:73,0.966475979879489)
--(axis cs:72,0.961599060572152)
--(axis cs:71,0.961599060572152)
--(axis cs:70,0.961599060572152)
--(axis cs:69,0.956012669613004)
--(axis cs:68,0.956012669613004)
--(axis cs:67,0.956012669613004)
--(axis cs:66,0.949613596349374)
--(axis cs:65,0.949613596349374)
--(axis cs:64,0.949613596349374)
--(axis cs:63,0.942283615520475)
--(axis cs:62,0.942283615520475)
--(axis cs:61,0.942283615520475)
--(axis cs:60,0.933887303001689)
--(axis cs:59,0.933887303001689)
--(axis cs:58,0.933887303001689)
--(axis cs:57,0.924269533793458)
--(axis cs:56,0.924269533793458)
--(axis cs:55,0.924269533793458)
--(axis cs:54,0.913252616029162)
--(axis cs:53,0.913252616029162)
--(axis cs:52,0.913252616029162)
--(axis cs:51,0.900633008051732)
--(axis cs:50,0.900633008051732)
--(axis cs:49,0.900633008051732)
--(axis cs:48,0.886177557905764)
--(axis cs:47,0.886177557905764)
--(axis cs:46,0.886177557905764)
--(axis cs:45,0.869619195768344)
--(axis cs:44,0.869619195768344)
--(axis cs:43,0.869619195768344)
--(axis cs:42,0.850651999734643)
--(axis cs:41,0.850651999734643)
--(axis cs:40,0.850651999734643)
--(axis cs:39,0.828925543796842)
--(axis cs:38,0.828925543796842)
--(axis cs:37,0.828925543796842)
--(axis cs:36,0.804038423593176)
--(axis cs:35,0.804038423593176)
--(axis cs:34,0.804038423593176)
--(axis cs:33,0.775530840312916)
--(axis cs:32,0.775530840312916)
--(axis cs:31,0.775530840312916)
--(axis cs:30,0.742876105742172)
--(axis cs:29,0.742876105742172)
--(axis cs:28,0.742876105742172)
--(axis cs:27,0.705470911503061)
--(axis cs:26,0.705470911503061)
--(axis cs:25,0.705470911503061)
--(axis cs:24,0.662624182706828)
--(axis cs:23,0.662624182706828)
--(axis cs:22,0.662624182706828)
--(axis cs:21,0.613544310088004)
--(axis cs:20,0.613544310088004)
--(axis cs:19,0.613544310088004)
--(axis cs:18,0.557324524728527)
--(axis cs:17,0.557324524728527)
--(axis cs:16,0.557324524728527)
--(axis cs:15,0.492926145164407)
--(axis cs:14,0.492926145164407)
--(axis cs:13,0.492926145164407)
--(axis cs:12,0.419159387359)
--(axis cs:11,0.419159387359)
--(axis cs:10,0.419159387359)
--(axis cs:9,0.334661383)
--(axis cs:8,0.334661383)
--(axis cs:7,0.334661383)
--(axis cs:6,0.237871)
--(axis cs:5,0.237871)
--(axis cs:4,0.237871)
--(axis cs:3,0.127)
--(axis cs:2,0.127)
--(axis cs:1,0.127)
--(axis cs:0,0)
--cycle;

\path [draw=black, fill=orange, opacity=0.5]
(axis cs:0,0)
--(axis cs:0,0)
--(axis cs:1,0.098)
--(axis cs:2,0.098)
--(axis cs:3,0.098)
--(axis cs:4,0.186396)
--(axis cs:5,0.186396)
--(axis cs:6,0.186396)
--(axis cs:7,0.266129192)
--(axis cs:8,0.266129192)
--(axis cs:9,0.266129192)
--(axis cs:10,0.338048531184)
--(axis cs:11,0.338048531184)
--(axis cs:12,0.338048531184)
--(axis cs:13,0.402919775127968)
--(axis cs:14,0.402919775127968)
--(axis cs:15,0.402919775127968)
--(axis cs:16,0.461433637165427)
--(axis cs:17,0.461433637165427)
--(axis cs:18,0.461433637165427)
--(axis cs:19,0.514213140723215)
--(axis cs:20,0.514213140723215)
--(axis cs:21,0.514213140723215)
--(axis cs:22,0.56182025293234)
--(axis cs:23,0.56182025293234)
--(axis cs:24,0.56182025293234)
--(axis cs:25,0.604761868144971)
--(axis cs:26,0.604761868144971)
--(axis cs:27,0.604761868144971)
--(axis cs:28,0.643495205066764)
--(axis cs:29,0.643495205066764)
--(axis cs:30,0.643495205066764)
--(axis cs:31,0.678432674970221)
--(axis cs:32,0.678432674970221)
--(axis cs:33,0.678432674970221)
--(axis cs:34,0.709946272823139)
--(axis cs:35,0.709946272823139)
--(axis cs:36,0.709946272823139)
--(axis cs:37,0.738371538086471)
--(axis cs:38,0.738371538086471)
--(axis cs:39,0.738371538086471)
--(axis cs:40,0.764011127353997)
--(axis cs:41,0.764011127353997)
--(axis cs:42,0.764011127353997)
--(axis cs:43,0.787138036873306)
--(axis cs:44,0.787138036873306)
--(axis cs:45,0.787138036873306)
--(axis cs:46,0.807998509259722)
--(axis cs:47,0.807998509259722)
--(axis cs:48,0.807998509259722)
--(axis cs:49,0.826814655352269)
--(axis cs:50,0.826814655352269)
--(axis cs:51,0.826814655352269)
--(axis cs:52,0.843786819127746)
--(axis cs:53,0.843786819127746)
--(axis cs:54,0.843786819127746)
--(axis cs:55,0.859095710853227)
--(axis cs:56,0.859095710853227)
--(axis cs:57,0.859095710853227)
--(axis cs:58,0.872904331189611)
--(axis cs:59,0.872904331189611)
--(axis cs:60,0.872904331189611)
--(axis cs:61,0.885359706733029)
--(axis cs:62,0.885359706733029)
--(axis cs:63,0.885359706733029)
--(axis cs:64,0.896594455473192)
--(axis cs:65,0.896594455473192)
--(axis cs:66,0.896594455473192)
--(axis cs:67,0.90672819883682)
--(axis cs:68,0.90672819883682)
--(axis cs:69,0.90672819883682)
--(axis cs:70,0.915868835350811)
--(axis cs:71,0.915868835350811)
--(axis cs:72,0.915868835350811)
--(axis cs:73,0.924113689486432)
--(axis cs:74,0.924113689486432)
--(axis cs:75,0.924113689486432)
--(axis cs:76,0.931550547916761)
--(axis cs:77,0.931550547916761)
--(axis cs:78,0.931550547916761)
--(axis cs:79,0.938258594220919)
--(axis cs:80,0.938258594220919)
--(axis cs:81,0.938258594220919)
--(axis cs:82,0.944309251987269)
--(axis cs:83,0.944309251987269)
--(axis cs:84,0.944309251987269)
--(axis cs:85,0.949766945292516)
--(axis cs:86,0.949766945292516)
--(axis cs:87,0.949766945292516)
--(axis cs:88,0.95468978465385)
--(axis cs:89,0.95468978465385)
--(axis cs:90,0.95468978465385)
--(axis cs:91,0.959130185757773)
--(axis cs:92,0.959130185757773)
--(axis cs:93,0.959130185757773)
--(axis cs:94,0.963135427553511)
--(axis cs:95,0.963135427553511)
--(axis cs:96,0.963135427553511)
--(axis cs:97,0.966748155653267)
--(axis cs:98,0.966748155653267)
--(axis cs:99,0.966748155653267)
--(axis cs:100,0.970006836399247)
--(axis cs:100,0.986006321417544)
--(axis cs:100,0.986006321417544)
--(axis cs:99,0.984134151267057)
--(axis cs:98,0.984134151267057)
--(axis cs:97,0.984134151267057)
--(axis cs:96,0.98201150937308)
--(axis cs:95,0.98201150937308)
--(axis cs:94,0.98201150937308)
--(axis cs:93,0.979604885910522)
--(axis cs:92,0.979604885910522)
--(axis cs:91,0.979604885910522)
--(axis cs:90,0.976876287880411)
--(axis cs:89,0.976876287880411)
--(axis cs:88,0.976876287880411)
--(axis cs:87,0.973782639320193)
--(axis cs:86,0.973782639320193)
--(axis cs:85,0.973782639320193)
--(axis cs:84,0.970275101270061)
--(axis cs:83,0.970275101270061)
--(axis cs:82,0.970275101270061)
--(axis cs:81,0.966298300759706)
--(axis cs:80,0.966298300759706)
--(axis cs:79,0.966298300759706)
--(axis cs:78,0.961789456643657)
--(axis cs:77,0.961789456643657)
--(axis cs:76,0.961789456643657)
--(axis cs:75,0.956677388484872)
--(axis cs:74,0.956677388484872)
--(axis cs:73,0.956677388484872)
--(axis cs:72,0.950881392839991)
--(axis cs:71,0.950881392839991)
--(axis cs:70,0.950881392839991)
--(axis cs:69,0.944309969206339)
--(axis cs:68,0.944309969206339)
--(axis cs:67,0.944309969206339)
--(axis cs:66,0.936859375517392)
--(axis cs:65,0.936859375517392)
--(axis cs:64,0.936859375517392)
--(axis cs:63,0.92841199038253)
--(axis cs:62,0.92841199038253)
--(axis cs:61,0.92841199038253)
--(axis cs:60,0.91883445621602)
--(axis cs:59,0.91883445621602)
--(axis cs:58,0.91883445621602)
--(axis cs:57,0.907975573941066)
--(axis cs:56,0.907975573941066)
--(axis cs:55,0.907975573941066)
--(axis cs:54,0.895663916032955)
--(axis cs:53,0.895663916032955)
--(axis cs:52,0.895663916032955)
--(axis cs:51,0.88170512021877)
--(axis cs:50,0.88170512021877)
--(axis cs:49,0.88170512021877)
--(axis cs:48,0.865878821109716)
--(axis cs:47,0.865878821109716)
--(axis cs:46,0.865878821109716)
--(axis cs:45,0.847935171326209)
--(axis cs:44,0.847935171326209)
--(axis cs:43,0.847935171326209)
--(axis cs:42,0.827590897195248)
--(axis cs:41,0.827590897195248)
--(axis cs:40,0.827590897195248)
--(axis cs:39,0.804524826751982)
--(axis cs:38,0.804524826751982)
--(axis cs:37,0.804524826751982)
--(axis cs:36,0.778372819446692)
--(axis cs:35,0.778372819446692)
--(axis cs:34,0.778372819446692)
--(axis cs:33,0.748722017513256)
--(axis cs:32,0.748722017513256)
--(axis cs:31,0.748722017513256)
--(axis cs:30,0.715104328246322)
--(axis cs:29,0.715104328246322)
--(axis cs:28,0.715104328246322)
--(axis cs:27,0.676989034292882)
--(axis cs:26,0.676989034292882)
--(axis cs:25,0.676989034292882)
--(axis cs:24,0.633774415298052)
--(axis cs:23,0.633774415298052)
--(axis cs:22,0.633774415298052)
--(axis cs:21,0.584778248637247)
--(axis cs:20,0.584778248637247)
--(axis cs:19,0.584778248637247)
--(axis cs:18,0.529227039271255)
--(axis cs:17,0.529227039271255)
--(axis cs:16,0.529227039271255)
--(axis cs:15,0.466243808697568)
--(axis cs:14,0.466243808697568)
--(axis cs:13,0.466243808697568)
--(axis cs:12,0.394834250224)
--(axis cs:11,0.394834250224)
--(axis cs:10,0.394834250224)
--(axis cs:9,0.313871032)
--(axis cs:8,0.313871032)
--(axis cs:7,0.313871032)
--(axis cs:6,0.222076)
--(axis cs:5,0.222076)
--(axis cs:4,0.222076)
--(axis cs:3,0.118)
--(axis cs:2,0.118)
--(axis cs:1,0.118)
--(axis cs:0,0)
--cycle;

\path [draw=black, fill=red, opacity=0.5]
(axis cs:0,0)
--(axis cs:0,0)
--(axis cs:1,0.099)
--(axis cs:2,0.099)
--(axis cs:3,0.099)
--(axis cs:4,0.188199)
--(axis cs:5,0.188199)
--(axis cs:6,0.188199)
--(axis cs:7,0.268567299)
--(axis cs:8,0.268567299)
--(axis cs:9,0.268567299)
--(axis cs:10,0.340979136399)
--(axis cs:11,0.340979136399)
--(axis cs:12,0.340979136399)
--(axis cs:13,0.406222201895499)
--(axis cs:14,0.406222201895499)
--(axis cs:15,0.406222201895499)
--(axis cs:16,0.465006203907845)
--(axis cs:17,0.465006203907845)
--(axis cs:18,0.465006203907845)
--(axis cs:19,0.517970589720968)
--(axis cs:20,0.517970589720968)
--(axis cs:21,0.517970589720968)
--(axis cs:22,0.565691501338592)
--(axis cs:23,0.565691501338592)
--(axis cs:24,0.565691501338592)
--(axis cs:25,0.608688042706071)
--(axis cs:26,0.608688042706071)
--(axis cs:27,0.608688042706071)
--(axis cs:28,0.64742792647817)
--(axis cs:29,0.64742792647817)
--(axis cs:30,0.64742792647817)
--(axis cs:31,0.682332561756832)
--(axis cs:32,0.682332561756832)
--(axis cs:33,0.682332561756832)
--(axis cs:34,0.713781638142905)
--(axis cs:35,0.713781638142905)
--(axis cs:36,0.713781638142905)
--(axis cs:37,0.742117255966758)
--(axis cs:38,0.742117255966758)
--(axis cs:39,0.742117255966758)
--(axis cs:40,0.767647647626049)
--(axis cs:41,0.767647647626049)
--(axis cs:42,0.767647647626049)
--(axis cs:43,0.79065053051107)
--(axis cs:44,0.79065053051107)
--(axis cs:45,0.79065053051107)
--(axis cs:46,0.811376127990474)
--(axis cs:47,0.811376127990474)
--(axis cs:48,0.811376127990474)
--(axis cs:49,0.830049891319417)
--(axis cs:50,0.830049891319417)
--(axis cs:51,0.830049891319417)
--(axis cs:52,0.846874952078795)
--(axis cs:53,0.846874952078795)
--(axis cs:54,0.846874952078795)
--(axis cs:55,0.862034331822994)
--(axis cs:56,0.862034331822994)
--(axis cs:57,0.862034331822994)
--(axis cs:58,0.875692932972517)
--(axis cs:59,0.875692932972517)
--(axis cs:60,0.875692932972517)
--(axis cs:61,0.887999332608238)
--(axis cs:62,0.887999332608238)
--(axis cs:63,0.887999332608238)
--(axis cs:64,0.899087398680023)
--(axis cs:65,0.899087398680023)
--(axis cs:66,0.899087398680023)
--(axis cs:67,0.9090777462107)
--(axis cs:68,0.9090777462107)
--(axis cs:69,0.9090777462107)
--(axis cs:70,0.918079049335841)
--(axis cs:71,0.918079049335841)
--(axis cs:72,0.918079049335841)
--(axis cs:73,0.926189223451593)
--(axis cs:74,0.926189223451593)
--(axis cs:75,0.926189223451593)
--(axis cs:76,0.933496490329885)
--(axis cs:77,0.933496490329885)
--(axis cs:78,0.933496490329885)
--(axis cs:79,0.940080337787227)
--(axis cs:80,0.940080337787227)
--(axis cs:81,0.940080337787227)
--(axis cs:82,0.946012384346291)
--(axis cs:83,0.946012384346291)
--(axis cs:84,0.946012384346291)
--(axis cs:85,0.951357158296008)
--(axis cs:86,0.951357158296008)
--(axis cs:87,0.951357158296008)
--(axis cs:88,0.956172799624703)
--(axis cs:89,0.956172799624703)
--(axis cs:90,0.956172799624703)
--(axis cs:91,0.960511692461858)
--(axis cs:92,0.960511692461858)
--(axis cs:93,0.960511692461858)
--(axis cs:94,0.964421034908134)
--(axis cs:95,0.964421034908134)
--(axis cs:96,0.964421034908134)
--(axis cs:97,0.967943352452229)
--(axis cs:98,0.967943352452229)
--(axis cs:99,0.967943352452229)
--(axis cs:100,0.971116960559458)
--(axis cs:100,0.980237497528474)
--(axis cs:100,0.980237497528474)
--(axis cs:99,0.977819862545986)
--(axis cs:98,0.977819862545986)
--(axis cs:97,0.977819862545986)
--(axis cs:96,0.975106467503913)
--(axis cs:95,0.975106467503913)
--(axis cs:94,0.975106467503913)
--(axis cs:93,0.972061130756356)
--(axis cs:92,0.972061130756356)
--(axis cs:91,0.972061130756356)
--(axis cs:90,0.968643244395461)
--(axis cs:89,0.968643244395461)
--(axis cs:88,0.968643244395461)
--(axis cs:87,0.964807232767072)
--(axis cs:86,0.964807232767072)
--(axis cs:85,0.964807232767072)
--(axis cs:84,0.960501944744188)
--(axis cs:83,0.960501944744188)
--(axis cs:82,0.960501944744188)
--(axis cs:81,0.955669971654532)
--(axis cs:80,0.955669971654532)
--(axis cs:79,0.955669971654532)
--(axis cs:78,0.950246881767152)
--(axis cs:77,0.950246881767152)
--(axis cs:76,0.950246881767152)
--(axis cs:75,0.944160361130361)
--(axis cs:74,0.944160361130361)
--(axis cs:73,0.944160361130361)
--(axis cs:72,0.937329249304558)
--(axis cs:71,0.937329249304558)
--(axis cs:70,0.937329249304558)
--(axis cs:69,0.929662457131939)
--(axis cs:68,0.929662457131939)
--(axis cs:67,0.929662457131939)
--(axis cs:66,0.921057752112165)
--(axis cs:65,0.921057752112165)
--(axis cs:64,0.921057752112165)
--(axis cs:63,0.911400395187616)
--(axis cs:62,0.911400395187616)
--(axis cs:61,0.911400395187616)
--(axis cs:60,0.900561610760511)
--(axis cs:59,0.900561610760511)
--(axis cs:58,0.900561610760511)
--(axis cs:57,0.888396869540417)
--(axis cs:56,0.888396869540417)
--(axis cs:55,0.888396869540417)
--(axis cs:54,0.874743961324822)
--(axis cs:53,0.874743961324822)
--(axis cs:52,0.874743961324822)
--(axis cs:51,0.859420832014391)
--(axis cs:50,0.859420832014391)
--(axis cs:49,0.859420832014391)
--(axis cs:48,0.842223156020641)
--(axis cs:47,0.842223156020641)
--(axis cs:46,0.842223156020641)
--(axis cs:45,0.822921611695444)
--(axis cs:44,0.822921611695444)
--(axis cs:43,0.822921611695444)
--(axis cs:42,0.801258823451677)
--(axis cs:41,0.801258823451677)
--(axis cs:40,0.801258823451677)
--(axis cs:39,0.776945929799862)
--(axis cs:38,0.776945929799862)
--(axis cs:37,0.776945929799862)
--(axis cs:36,0.749658731537443)
--(axis cs:35,0.749658731537443)
--(axis cs:34,0.749658731537443)
--(axis cs:33,0.719033368728892)
--(axis cs:32,0.719033368728892)
--(axis cs:31,0.719033368728892)
--(axis cs:30,0.684661468831529)
--(axis cs:29,0.684661468831529)
--(axis cs:28,0.684661468831529)
--(axis cs:27,0.646084701269954)
--(axis cs:26,0.646084701269954)
--(axis cs:25,0.646084701269954)
--(axis cs:24,0.602788665847311)
--(axis cs:23,0.602788665847311)
--(axis cs:22,0.602788665847311)
--(axis cs:21,0.554196033498665)
--(axis cs:20,0.554196033498665)
--(axis cs:19,0.554196033498665)
--(axis cs:18,0.499658847922183)
--(axis cs:17,0.499658847922183)
--(axis cs:16,0.499658847922183)
--(axis cs:15,0.438449885434549)
--(axis cs:14,0.438449885434549)
--(axis cs:13,0.438449885434549)
--(axis cs:12,0.369752957839)
--(axis cs:11,0.369752957839)
--(axis cs:10,0.369752957839)
--(axis cs:9,0.292652029)
--(axis cs:8,0.292652029)
--(axis cs:7,0.292652029)
--(axis cs:6,0.206119)
--(axis cs:5,0.206119)
--(axis cs:4,0.206119)
--(axis cs:3,0.109)
--(axis cs:2,0.109)
--(axis cs:1,0.109)
--(axis cs:0,0)
--cycle;

\addplot [semithick, black, forget plot]
table {%
0 0
1 0.1
2 0.1
3 0.1
4 0.19
5 0.19
6 0.19
7 0.271
8 0.271
9 0.271
10 0.3439
11 0.3439
12 0.3439
13 0.40951
14 0.40951
15 0.40951
16 0.468559
17 0.468559
18 0.468559
19 0.5217031
20 0.5217031
21 0.5217031
22 0.56953279
23 0.56953279
24 0.56953279
25 0.612579511
26 0.612579511
27 0.612579511
28 0.6513215599
29 0.6513215599
30 0.6513215599
31 0.68618940391
32 0.68618940391
33 0.68618940391
34 0.717570463519
35 0.717570463519
36 0.717570463519
37 0.7458134171671
38 0.7458134171671
39 0.7458134171671
40 0.77123207545039
41 0.77123207545039
42 0.77123207545039
43 0.794108867905351
44 0.794108867905351
45 0.794108867905351
46 0.814697981114816
47 0.814697981114816
48 0.814697981114816
49 0.833228183003334
50 0.833228183003334
51 0.833228183003334
52 0.849905364703001
53 0.849905364703001
54 0.849905364703001
55 0.864914828232701
56 0.864914828232701
57 0.864914828232701
58 0.878423345409431
59 0.878423345409431
60 0.878423345409431
61 0.890581010868488
62 0.890581010868488
63 0.890581010868488
64 0.901522909781639
65 0.901522909781639
66 0.901522909781639
67 0.911370618803475
68 0.911370618803475
69 0.911370618803475
70 0.920233556923127
71 0.920233556923127
72 0.920233556923127
73 0.928210201230815
74 0.928210201230815
75 0.928210201230815
76 0.935389181107733
77 0.935389181107733
78 0.935389181107733
79 0.94185026299696
80 0.94185026299696
81 0.94185026299696
82 0.947665236697264
83 0.947665236697264
84 0.947665236697264
85 0.952898713027538
86 0.952898713027538
87 0.952898713027538
88 0.957608841724784
89 0.957608841724784
90 0.957608841724784
91 0.961847957552305
92 0.961847957552305
93 0.961847957552305
94 0.965663161797075
95 0.965663161797075
96 0.965663161797075
97 0.969096845617367
98 0.969096845617367
99 0.969096845617367
100 0.972187161055631
};
\addplot [semithick, black, dashed, forget plot]
table {%
-5 0.25
105 0.25
};
\addplot [semithick, black, dashed, forget plot]
table {%
-5 1
105 1
};
\end{axis}
\end{tikzpicture}
\caption{Hitting probability ranges for increasing perturbation levels.
Red, orange and yellow plots correspond to $\epsilon \in \{0.01,0.02,0.03\}$.}
\label{fig:sa}
\end{figure}
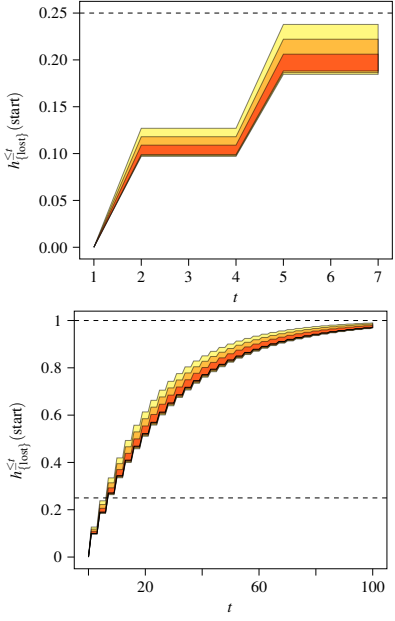

\begin{example}\label{ex:toy2}
Consider an imprecise probabilistic version of the model $M$ discussed in Example \ref{ex:toy}. The same, precise, transition probabilities are kept as in Figure \ref{fig:4states} apart from $P(S_{t+1}|s_t=\mathrm{try})$. For a sensitivity analysis parametrized by $\epsilon \in [0,1]$, we replace such (conditional) PMF with the CS induced by the linear constraint:
\begin{equation}
P(\mathrm{delivered}|\mathrm{try})\in[(1-\epsilon)0.9-\epsilon,(1-\epsilon)0.9+\epsilon]\,.
\end{equation}
with the impossible transitions remaining impossible.\footnote{As only two states are possible, the bounds on the probability of $\{\mathrm{lost}\}$ can be also induced by $\epsilon$-contamination. Deterministic PMFs are unaffected by the contamination.} This makes the model a IMC to be used to answer IPCTL queries, such as \emph{``the upper probability to lose a message within seven time steps is less than or equal to $0.25$''}. Having this formula for the upper probability  satisfied ensures that any MC consistent with the IMC would satisfy the analogous formula in Example \ref{ex:toy}, thus providing the desired sensitivity analysis. Figure \ref{fig:sa} shows the corresponding (upper) hitting probabilities for increasing values of $\epsilon$ computed by means of the recursion in Equation \eqref{eq:hitting_dual_imprecise} and two different time horizons. Even for the higher perturbation level we consider ($\epsilon=0.03$) within seven time steps the hitting probability remains under the threshold level. In the limit of longer chains, for this model, both bounds converge to one.  
\end{example}

\subsection{IPRCTL Syntax}\label{sec:iprctl_syntax}
We are now in the condition of extending IPCTL syntax in order to cope with IMRMs. We call the corresponding language IPRCTL. The IPRCTL syntax is obtained by augmenting the IPCTL syntax in Section \ref{sec:ipctlsyntax} with
the PRCTL path formulae in Equation \eqref{eq:prctl_syntax2}, while expression $E_{\nabla{r}}\phi$ in 
Equation \eqref{eq:prctl_syntax1}
is replaced by:
\begin{equation}
    \underline{E}_{\nabla{r}}\phi \mid \overline{E}_{\nabla{r}}\phi\,,
\end{equation}
whose informal reading is as for PRCTL in Section \ref{sec:prctlsyntax}.

\subsection{IPRCTL Semantics}\label{sec:iprctl_semantics}
The IPRCTL semantics is obtained extending the IPCTL semantics with respective satisfiability conditions for expected reward and reward-bounded probabilities.

\paragraph{Expected Cumulative Reward.}
Given an IMRM $(\mathcal{M},rew)$ and state $s\in \mathcal{S}$, the analogous of Equation \eqref{eq:exprewformula} corresponds to condition:
\begin{equation}\label{eq:iecr}
(\mathcal{M},rew),s\models \overline{E}_{\nabla{r}}\phi \,\, \mathrm{iff}
\,\, \overline{E}[Rew_{\Phi}^{\leq t}](s) \nabla{r}\,,
\end{equation}
and analogously for $\underline{E}_{\nabla{r}}$ and $\underline{E}[Rew_\Phi^{\leq t}]$.
Those lower and upper expected cumulative reward arrays $\underline{E}[Rew_{\Phi}^{\leq t}]$ and 
$\overline{E}[Rew_{\Phi}^{\leq t}]$ represent the lower and upper bounds of the precise expectations in Equation \eqref{eq:expcumrew} with respect to the CSs in the specification of $\mathcal{M}$ (see Section \ref{sec:imc}).

Exactly as for the derivation of Equation \eqref{eq:cond_lower_hitting}, we rely on the results by \cite{TJoensKBC19} to achieve an imprecise-probabilistic version of the recursion in Equation \eqref{eq:recursion_expcumrew}. This simply corresponds to:
\begin{equation}\label{ref:upperExpectedCumulativeReward}
\overline{E}[Rew_{\Phi}^{\leq \tau}]:= rew + \mathbb{I}_{\Phi^{c}}  \overline{\mathcal{T}} \,\overline{E}[Rew_{\Phi}^{\leq \tau-1}]\,,
\end{equation}
for each $\tau=1,\ldots,t$, and analogously for the lower expectation, with initialization for both cases:
\begin{equation}\label{ref:upperExpectedCumulativeRewardInit}
\underline{E}[Rew_{\Phi}^{\leq 0}]= 
\overline{E}[Rew_{\Phi}^{\leq 0}]= rew\,.
\end{equation}
The recursion in Equation \eqref{ref:upperExpectedCumulativeReward}, exactly as that in Equation \eqref{eq:hitting_dual_imprecise}, requires $t\in \mathbb{N}$ iterative applications of the non-linear transition operator in Equation \eqref{eq:impreciseT}. As for Equation \eqref{eq:cond_lower_hitting}, each further iterative application of \eqref{ref:upperExpectedCumulativeReward} requires to solve $|\mathcal{S}|$ linear programming tasks. Also in this case, hence, shifting to imprecise probabilities does not increases the overall computational complexity with respect to $|\mathcal{S}|$. Notice that, the complexity with respect to the nesting depth of $\phi$ results to be exponential also in this case, but the same practical considerations stated for Equations \eqref{eq:cond_lower_hitting} hold.

\paragraph{Bounded-Reward Probability.} Given IMRM $(\mathcal{M},rew)$ and state $s\in \mathcal{S}$, the following condition holds:
\begin{equation}
(\mathcal{M},rew),s \models \overline{P}_{\nabla{b}}\delta \,\,\mathrm{iff}\,\,\overline{P}(s\models \phi_1\bigcup_{\leq r}\phi_2)
\nabla{r}
\,,
\end{equation}
where the event on the right-hand side is as in Equation \eqref{eq:bounded_rew_semantics} and an analogous condition holds for the lower probability. By a discussion similar to that in Section \ref{sec:prctlsemantics}, we obtain a recursive relation analogous to Equation \eqref{ref:RecursionBoundedReward} 
for the upper probabilities, denoted here as $\overline{x}_{\Phi_2|\Phi_1}^{\leq \tau,\rho}$, by simply replacing the linear operator with its non-linear, upper, version, i.e.,
\begin{equation}\label{eq:Bounded_imprecise}
\overline{x}^{\leq \tau, \rho}_{\Phi_2|\Phi_1}:= \mathbb{I}_{\mathcal{S}^{rew}_{\rho}} \left(\mathbb{I}_{\Phi_2} + \mathbb{I}_{\Phi_1\setminus\Phi_2}\cdot \overline{\mathcal{T}} \overline{\chi}^{\leq \tau-1,\rho} \right)\,,
\end{equation}
where $\overline{\chi}^{\leq \tau-1,\rho}$ is obtained as in Equation \eqref{eq:chi} but from the upper probabilities for the same time step. An analogous derivation holds for the lower bound. Both the initializations are as in the precise case in Equation \eqref{eq:init_bounded_prob}.
Notice that, Equation \eqref{eq:Bounded_imprecise} can be obtained from Equation \eqref{eq:cond_lower_hitting} by simply including the indicator vector $\mathbb{I}_{\mathcal{S}_{\rho}^{rew}}$, which blocks the recursion for each $s\in \mathcal{S}: rew(s)> \rho$, and by replacing the upper hitting probability with the function $\overline{\chi}^{\leq \tau-1,\rho}$. Note that $\overline{\chi}^{\leq \tau-1,\rho}$ coincides with $\overline{x}^{\leq \tau,\rho-rew(s)}_{\Phi_2|\Phi_1}$. For each further iteration hence, the reward threshold $\rho$ is reduced, for each $s\in \mathcal{S}$, of a value $rew(s)$. 
The recursion is eventually blocked when all the reached states $s'\in \mathcal{S}$ are either such that $s'\in \Phi_2$ or such that $rew(s')> \rho$. Also in this case this always happens for a finite time horizon $t\in \mathbb{N}$ because the reward threshold $\rho$ is reduced at each further iteration.

Concerning computational complexity, since Equation \eqref{eq:Bounded_imprecise} also requires an application of the dual imprecise transition operator $\mathcal{T}^{\dagger}$ for each further recursive step, the same considerations stated for Equation \eqref{eq:cond_lower_hitting} hold. The overall computational complexity of \eqref{eq:Bounded_imprecise} is therefore polynomial with respect to $|\mathcal{S}|\cdot t$, where $t$ is the total number of iterations occured until any further iteration is blocked.

\section{A Case Study on IPRCTL}\label{sec:case_study}
As a very first IPRCTL application we perform a sensitivity analysis in the MRM originally proposed by \citet{MRMGeriatric}. In that paper expected cumulative rewards are used to estimate the cost of annual recovery of geriatric patients. Let us briefly describe their model and report the results of our IPRCTL-based sensitivity analysis.\footnote{Code available at \href{https://github.com/IDSIA-papers/2021-ISIPTA-IPRCTL}{github.com/IDSIA-papers/2021-ISIPTA-IPRCTL}.}

\begin{figure}[htp!]
\centering
\begin{tikzpicture}[style=thick,shorten >=0.1pt,->,every 
path/.style={-latex},scale=1.0]
\tikzstyle{vertex}=[align=center,circle,text width=10mm,draw,fill=white,minimum size=12pt,inner sep=1pt]
\centering
\node[vertex]  (A)  at  (0, 0) {\footnotesize A (100\pounds)};  
\node[vertex]  (L)  at  (4, 0) {\footnotesize L (50\pounds)};
\node[vertex]  (D)  at  (2,1) {\footnotesize D (0\pounds)}; 
\path[->](A)  edge [bend right] node[below]  {$\nu$}  (L);
\path[->](D)  edge [loop below] node[right]  {\phantom{a}$1$} ();
\path[->](L)  edge [loop above] node[above]  {$1-\delta$} ();
\path[->](A)  edge [loop above] node[above]  {$1-\nu-\gamma$\phantom{a}} ();
\path[->](L)  edge [] node[above]  {\phantom{a} $\delta$}  (D);
\path[->](A)  edge [] node[above]  {$\gamma$}  (D);
\end{tikzpicture}
\caption{Transitions in a three-state MRM.\label{fig:geriatric1}}
\end{figure}
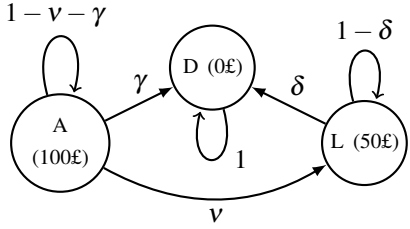

In the considered geriatric departments, there are two kinds of recovery: short-term recoveries for acute cares have a daily cost estimated as \pounds 100, while long-term recoveries cost \pounds 50. From a cumulative perspective, long-term recoveries are more expensive, since those patients typically remain in the hospitals for longer periods. The scenario can be naturally described as a PMC $M$ whose three states are in one-to-one correspondence with the singletons of the three atomic propositions: $A$ (acute care), $L$ (long stay), and $D$ (discharge or death). The first two states represent short and long-term recoveries, while the latter represents the end of a recovery. $D$ should be regarded as an absorbing state and a parametrized version of the transition matrix for this model is in Figure \ref{fig:geriatric1}. The parameters have the following interpretation: the \emph{conversion rate} $\nu$ corresponds to the probability of passing from a short-term to a long-term recovery, while the \emph{dismissing rates} $\gamma$ and $\delta$ correspond to the probability of being discharged/die, respectively, in a short- and long-term recovery. Rates $\gamma$, $\nu$ and $\delta$ vary depending on the patient and disease. An assessment of these parameters for different departments is in Table \ref{tab:parameters}.

\begin{table}[htp!]
\begin{center}
\begin{tabular}{crrr}
\hline
Rate  (\%) & Dep.1 & Dep.2 & Dep.3 \\
\hline
$\gamma$& 1.750  & 3.540 & 2.810 \\
$\nu$ & 0.031 & 0.187 & 0.149 \\
$\delta$ & 0.120 & 0.130 & 0.180\\
\hline
\end{tabular}
\end{center}
\caption{Conversion and dismissing rates.}\label{tab:parameters}
\end{table}

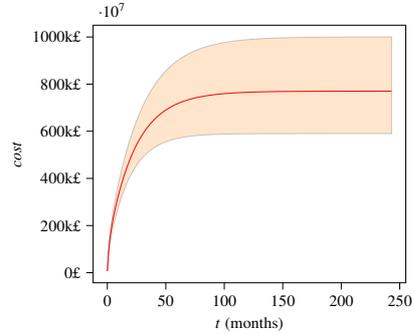
\begin{figure}[htp!]
\centering
\begin{tikzpicture}[scale=.6]
\begin{axis}[
tick align=outside,
tick pos=left,
x grid style={white!69.0196078431373!black},
xlabel={$t$ (months)},
xmin=-12.15, xmax=255.15,
xtick style={color=black},
y grid style={white!69.0196078431373!black},
ylabel={$cost$},
ymin=-429208.34341023, ymax=10500024.6616148,
ytick style={color=black},
ytick={0,2000000,4000000,6000000,8000000,10000000},
yticklabels={0\pounds,200k\pounds,400k\pounds,600k\pounds,800k\pounds,1000k\pounds}]
\path [draw=black, fill=orange, opacity=0.2]
(axis cs:0,68047.525)
--(axis cs:0,67574.975)
--(axis cs:1,767800.397083855)
--(axis cs:2,1172059.95864657)
--(axis cs:3,1466424.50918359)
--(axis cs:4,1714987.36946007)
--(axis cs:5,1940289.40768206)
--(axis cs:6,2150316.80073451)
--(axis cs:7,2348144.78007553)
--(axis cs:8,2535178.26133969)
--(axis cs:9,2712241.50834154)
--(axis cs:10,2879945.19747507)
--(axis cs:11,3038810.69562559)
--(axis cs:12,3189312.7310498)
--(axis cs:13,3331894.60258244)
--(axis cs:14,3466974.11798598)
--(axis cs:15,3594946.37444187)
--(axis cs:16,3716185.43692929)
--(axis cs:17,3831045.60722312)
--(axis cs:18,3939862.51827101)
--(axis cs:19,4042954.1347611)
--(axis cs:20,4140621.68887577)
--(axis cs:21,4233150.56272347)
--(axis cs:22,4320811.12296904)
--(axis cs:23,4403859.51108953)
--(axis cs:24,4482538.39189645)
--(axis cs:25,4557077.66262333)
--(axis cs:26,4627695.12468783)
--(axis cs:27,4694597.12010386)
--(axis cs:28,4757979.13440754)
--(axis cs:29,4818026.36786023)
--(axis cs:30,4874914.27659802)
--(axis cs:31,4928809.08530903)
--(axis cs:32,4979868.27293685)
--(axis cs:33,5028241.03282904)
--(axis cs:34,5074068.70867542)
--(axis cs:35,5117485.20750992)
--(axis cs:36,5158617.39098275)
--(axis cs:37,5197585.44604612)
--(axis cs:38,5234503.23613684)
--(axis cs:39,5269478.63388161)
--(axis cs:40,5302613.83629756)
--(axis cs:41,5334005.66340866)
--(axis cs:42,5363745.84115083)
--(axis cs:43,5391921.26939226)
--(axis cs:44,5418614.27585206)
--(axis cs:45,5443902.85665936)
--(axis cs:46,5467860.90425545)
--(axis cs:47,5490558.42330529)
--(axis cs:48,5512061.73524886)
--(axis cs:49,5532433.67209037)
--(axis cs:50,5551733.75999142)
--(axis cs:51,5570018.39320456)
--(axis cs:52,5587340.99885554)
--(axis cs:53,5603752.19305568)
--(axis cs:54,5619299.9288006)
--(axis cs:55,5634029.63608733)
--(axis cs:56,5647984.35465937)
--(axis cs:57,5661204.85976746)
--(axis cs:58,5673729.78131358)
--(axis cs:59,5685595.71672632)
--(axis cs:60,5696837.33789744)
--(axis cs:61,5707487.49249201)
--(axis cs:62,5717577.29992829)
--(axis cs:63,5727136.24230769)
--(axis cs:64,5736192.25056056)
--(axis cs:65,5744771.78605945)
--(axis cs:66,5752899.91793847)
--(axis cs:67,5760600.39634447)
--(axis cs:68,5767895.72183424)
--(axis cs:69,5774807.21112045)
--(axis cs:70,5781355.05935838)
--(axis cs:71,5787558.39915557)
--(axis cs:72,5793435.35647668)
--(axis cs:73,5799003.10360699)
--(axis cs:74,5804277.90932929)
--(axis cs:75,5809275.18646073)
--(axis cs:76,5814009.53688862)
--(axis cs:77,5818494.79423669)
--(axis cs:78,5822744.06428647)
--(axis cs:79,5826769.76327199)
--(axis cs:80,5830583.65415964)
--(axis cs:81,5834196.88101919)
--(axis cs:82,5837620.00158639)
--(axis cs:83,5840863.01811234)
--(axis cs:84,5843935.40658981)
--(axis cs:85,5846846.14444171)
--(axis cs:86,5849603.73675293)
--(axis cs:87,5852216.24112184)
--(axis cs:88,5854691.29120443)
--(axis cs:89,5857036.11901952)
--(axis cs:90,5859257.57608059)
--(axis cs:91,5861362.15341561)
--(axis cs:92,5863356.00053365)
--(axis cs:93,5865244.94339357)
--(axis cs:94,5867034.50142725)
--(axis cs:95,5868729.90366725)
--(axis cs:96,5870336.10402581)
--(axis cs:97,5871857.79577011)
--(axis cs:98,5873299.42523577)
--(axis cs:99,5874665.20481895)
--(axis cs:100,5875959.12528482)
--(axis cs:101,5877184.96742849)
--(axis cs:102,5878346.3131223)
--(axis cs:103,5879446.55578202)
--(axis cs:104,5880488.91028225)
--(axis cs:105,5881476.42235019)
--(axis cs:106,5882411.9774652)
--(axis cs:107,5883298.30929002)
--(axis cs:108,5884138.0076585)
--(axis cs:109,5884933.52614298)
--(axis cs:110,5885687.18922356)
--(axis cs:111,5886401.19908016)
--(axis cs:112,5887077.64202724)
--(axis cs:113,5887718.49460989)
--(axis cs:114,5888325.62937931)
--(axis cs:115,5888900.82036427)
--(axis cs:116,5889445.74825484)
--(axis cs:117,5889962.00531323)
--(axis cs:118,5890451.10002642)
--(axis cs:119,5890914.46151384)
--(axis cs:120,5891353.44370325)
--(axis cs:121,5891769.32928685)
--(axis cs:122,5892163.33346931)
--(axis cs:123,5892536.60751851)
--(axis cs:124,5892890.24212953)
--(axis cs:125,5893225.27061164)
--(axis cs:126,5893542.67190757)
--(axis cs:127,5893843.37345396)
--(axis cs:128,5894128.25389133)
--(axis cs:129,5894398.14563138)
--(axis cs:130,5894653.83728934)
--(axis cs:131,5894896.07598822)
--(axis cs:132,5895125.56954192)
--(axis cs:133,5895342.98852341)
--(axis cs:134,5895548.96822415)
--(axis cs:135,5895744.11051037)
--(axis cs:136,5895928.98558174)
--(axis cs:137,5896104.13363741)
--(axis cs:138,5896270.06645452)
--(axis cs:139,5896427.26888357)
--(axis cs:140,5896576.20026513)
--(axis cs:141,5896717.29577204)
--(axis cs:142,5896850.96768095)
--(axis cs:143,5896977.60657699)
--(axis cs:144,5897097.58249506)
--(axis cs:145,5897211.24600102)
--(axis cs:146,5897318.92921609)
--(axis cs:147,5897420.94678724)
--(axis cs:148,5897517.59680663)
--(axis cs:149,5897609.16168258)
--(axis cs:150,5897695.9089648)
--(axis cs:151,5897778.09212613)
--(axis cs:152,5897855.95130319)
--(axis cs:153,5897929.71399806)
--(axis cs:154,5897999.59574303)
--(axis cs:155,5898065.80073037)
--(axis cs:156,5898128.52240897)
--(axis cs:157,5898187.94404961)
--(axis cs:158,5898244.23928048)
--(axis cs:159,5898297.57259447)
--(axis cs:160,5898348.09982985)
--(axis cs:161,5898395.96862563)
--(axis cs:162,5898441.31885294)
--(axis cs:163,5898484.2830237)
--(axis cs:164,5898524.98667788)
--(axis cs:165,5898563.5487503)
--(axis cs:166,5898600.08191811)
--(axis cs:167,5898634.69293008)
--(axis cs:168,5898667.48291852)
--(axis cs:169,5898698.54769475)
--(axis cs:170,5898727.97802905)
--(axis cs:171,5898755.85991595)
--(axis cs:172,5898782.27482546)
--(axis cs:173,5898807.2999411)
--(axis cs:174,5898831.00838548)
--(axis cs:175,5898853.46943395)
--(axis cs:176,5898874.74871696)
--(axis cs:177,5898894.90841193)
--(axis cs:178,5898914.00742485)
--(axis cs:179,5898932.10156245)
--(axis cs:180,5898949.24369521)
--(axis cs:181,5898965.48391193)
--(axis cs:182,5898980.869666)
--(axis cs:183,5898995.44591411)
--(axis cs:184,5899009.2552476)
--(axis cs:185,5899022.33801692)
--(axis cs:186,5899034.73244951)
--(axis cs:187,5899046.47476152)
--(axis cs:188,5899057.5992636)
--(axis cs:189,5899068.13846121)
--(axis cs:190,5899078.12314955)
--(axis cs:191,5899087.58250356)
--(axis cs:192,5899096.5441632)
--(axis cs:193,5899105.03431413)
--(axis cs:194,5899113.07776433)
--(axis cs:195,5899120.69801649)
--(axis cs:196,5899127.91733676)
--(axis cs:197,5899134.75681977)
--(axis cs:198,5899141.23645026)
--(axis cs:199,5899147.37516152)
--(axis cs:200,5899153.19089068)
--(axis cs:201,5899158.70063109)
--(axis cs:202,5899163.92048207)
--(axis cs:203,5899168.86569584)
--(axis cs:204,5899173.55072217)
--(axis cs:205,5899177.98925056)
--(axis cs:206,5899182.19425025)
--(axis cs:207,5899186.17800811)
--(axis cs:208,5899189.95216456)
--(axis cs:209,5899193.52774758)
--(axis cs:210,5899196.9152049)
--(axis cs:211,5899200.12443456)
--(axis cs:212,5899203.16481383)
--(axis cs:213,5899206.04522661)
--(axis cs:214,5899208.77408937)
--(axis cs:215,5899211.35937577)
--(axis cs:216,5899213.80863993)
--(axis cs:217,5899216.12903852)
--(axis cs:218,5899218.32735167)
--(axis cs:219,5899220.41000279)
--(axis cs:220,5899222.38307732)
--(axis cs:221,5899224.25234052)
--(axis cs:222,5899226.02325432)
--(axis cs:223,5899227.70099328)
--(axis cs:224,5899229.29045969)
--(axis cs:225,5899230.79629792)
--(axis cs:226,5899232.22290799)
--(axis cs:227,5899233.57445841)
--(axis cs:228,5899234.85489837)
--(axis cs:229,5899236.06796927)
--(axis cs:230,5899237.21721566)
--(axis cs:231,5899238.30599561)
--(axis cs:232,5899239.33749051)
--(axis cs:233,5899240.31471433)
--(axis cs:234,5899241.24052251)
--(axis cs:235,5899242.11762023)
--(axis cs:236,5899242.94857033)
--(axis cs:237,5899243.73580083)
--(axis cs:238,5899244.48161199)
--(axis cs:239,5899245.18818304)
--(axis cs:240,5899245.85757857)
--(axis cs:241,5899246.49175453)
--(axis cs:242,5899247.09256396)
--(axis cs:243,5899247.66176241)
--(axis cs:243,10003241.3432046)
--(axis cs:243,10003241.3432046)
--(axis cs:242,10003190.5986087)
--(axis cs:241,10003137.9927895)
--(axis cs:240,10003083.4574807)
--(axis cs:239,10003026.9219117)
--(axis cs:238,10002968.3127167)
--(axis cs:237,10002907.5538385)
--(axis cs:236,10002844.5664307)
--(axis cs:235,10002779.2687546)
--(axis cs:234,10002711.5760735)
--(axis cs:233,10002641.400543)
--(axis cs:232,10002568.6510963)
--(axis cs:231,10002493.2333268)
--(axis cs:230,10002415.049365)
--(axis cs:229,10002333.9977518)
--(axis cs:228,10002249.9733067)
--(axis cs:227,10002162.8669915)
--(axis cs:226,10002072.5657685)
--(axis cs:225,10001978.9524542)
--(axis cs:224,10001881.9055666)
--(axis cs:223,10001781.2991685)
--(axis cs:222,10001677.0027031)
--(axis cs:221,10001568.8808251)
--(axis cs:220,10001456.7932252)
--(axis cs:219,10001340.5944476)
--(axis cs:218,10001220.1337014)
--(axis cs:217,10001095.2546651)
--(axis cs:216,10000965.7952834)
--(axis cs:215,10000831.5875574)
--(axis cs:214,10000692.457326)
--(axis cs:213,10000548.2240402)
--(axis cs:212,10000398.700529)
--(axis cs:211,10000243.6927562)
--(axis cs:210,10000082.9995685)
--(axis cs:209,9999916.41243482)
--(axis cs:208,9999743.71517557)
--(axis cs:207,9999564.68368195)
--(axis cs:206,9999379.0856253)
--(axis cs:205,9999186.68015554)
--(axis cs:204,9998987.21758862)
--(axis cs:203,9998780.43908252)
--(axis cs:202,9998566.07630135)
--(axis cs:201,9998343.85106713)
--(axis cs:200,9998113.47499878)
--(axis cs:199,9997874.64913792)
--(axis cs:198,9997627.06356089)
--(axis cs:197,9997370.39697658)
--(axis cs:196,9997104.31630947)
--(axis cs:195,9996828.47626742)
--(axis cs:194,9996542.51889358)
--(axis cs:193,9996246.07310186)
--(axis cs:192,9995938.75419536)
--(axis cs:191,9995620.16336722)
--(axis cs:190,9995289.88718298)
--(axis cs:189,9994947.49704416)
--(axis cs:188,9994592.54863203)
--(axis cs:187,9994224.58133102)
--(axis cs:186,9993843.11763099)
--(axis cs:185,9993447.66250755)
--(axis cs:184,9993037.70277969)
--(axis cs:183,9992612.7064438)
--(axis cs:182,9992172.12198332)
--(axis cs:181,9991715.37765303)
--(axis cs:180,9991241.88073706)
--(axis cs:179,9990751.01677979)
--(axis cs:178,9990242.14878842)
--(axis cs:177,9989714.61640638)
--(axis cs:176,9989167.73505635)
--(axis cs:175,9988600.79505194)
--(axis cs:174,9988013.06067666)
--(axis cs:173,9987403.76922925)
--(axis cs:172,9986772.13003389)
--(axis cs:171,9986117.32341416)
--(axis cs:170,9985438.49962931)
--(axis cs:169,9984734.77777162)
--(axis cs:168,9984005.24462318)
--(axis cs:167,9983248.95347084)
--(axis cs:166,9982464.92287768)
--(axis cs:165,9981652.13540938)
--(axis cs:164,9980809.5363139)
--(axis cs:163,9979936.03215273)
--(axis cs:162,9979030.48938193)
--(axis cs:161,9978091.73288118)
--(axis cs:160,9977118.54442878)
--(axis cs:159,9976109.66112076)
--(axis cs:158,9975063.77373206)
--(axis cs:157,9973979.52501753)
--(axis cs:156,9972855.50795061)
--(axis cs:155,9971690.26389749)
--(axis cs:154,9970482.28072418)
--(axis cs:153,9969229.99083428)
--(axis cs:152,9967931.76913467)
--(axis cs:151,9966585.93092666)
--(axis cs:150,9965190.72971974)
--(axis cs:149,9963744.35496519)
--(axis cs:148,9962244.9297065)
--(axis cs:147,9960690.50814367)
--(axis cs:146,9959079.07310816)
--(axis cs:145,9957408.53344521)
--(axis cs:144,9955676.72130012)
--(axis cs:143,9953881.38930508)
--(axis cs:142,9952020.20766276)
--(axis cs:141,9950090.7611229)
--(axis cs:140,9948090.54584808)
--(axis cs:139,9946016.96616451)
--(axis cs:138,9943867.33119357)
--(axis cs:137,9941638.85135993)
--(axis cs:136,9939328.63477151)
--(axis cs:135,9936933.68346666)
--(axis cs:134,9934450.88952371)
--(axis cs:133,9931877.03102786)
--(axis cs:132,9929208.76789009)
--(axis cs:131,9926442.63751274)
--(axis cs:130,9923575.05029608)
--(axis cs:129,9920602.28498016)
--(axis cs:128,9917520.48381569)
--(axis cs:127,9914325.64755784)
--(axis cs:126,9911013.63027648)
--(axis cs:125,9907580.13397597)
--(axis cs:124,9904020.7030177)
--(axis cs:123,9900330.71833801)
--(axis cs:122,9896505.39145401)
--(axis cs:121,9892539.75824964)
--(axis cs:120,9888428.67253369)
--(axis cs:119,9884166.79936166)
--(axis cs:118,9879748.60811255)
--(axis cs:117,9875168.36531185)
--(axis cs:116,9870420.12719121)
--(axis cs:115,9865497.73197518)
--(axis cs:114,9860394.79188519)
--(axis cs:113,9855104.68485002)
--(axis cs:112,9849620.54591249)
--(axis cs:111,9843935.25832071)
--(axis cs:110,9838041.44429277)
--(axis cs:109,9831931.45544257)
--(axis cs:108,9825597.36285454)
--(axis cs:107,9819030.94679432)
--(axis cs:106,9812223.68604203)
--(axis cs:105,9805166.74683429)
--(axis cs:104,9797850.97140067)
--(axis cs:103,9790266.86607973)
--(axis cs:102,9782404.58899905)
--(axis cs:101,9774253.93730347)
--(axis cs:100,9765804.33391495)
--(axis cs:99,9757044.81380663)
--(axis cs:98,9747964.00977357)
--(axis cs:97,9738550.13768161)
--(axis cs:96,9728790.98117507)
--(axis cs:95,9718673.87582365)
--(axis cs:94,9708185.69268783)
--(axis cs:93,9697312.82128149)
--(axis cs:92,9686041.15190958)
--(axis cs:91,9674356.05735807)
--(axis cs:90,9662242.3739122)
--(axis cs:89,9649684.3816786)
--(axis cs:88,9636665.78418563)
--(axis cs:87,9623169.68723546)
--(axis cs:86,9609178.57698056)
--(axis cs:85,9594674.29719601)
--(axis cs:84,9579638.0257182)
--(axis cs:83,9564050.25001941)
--(axis cs:82,9547890.74188636)
--(axis cs:81,9531138.53117025)
--(axis cs:80,9513771.87857374)
--(axis cs:79,9495768.24743998)
--(axis cs:78,9477104.27450685)
--(axis cs:77,9457755.73958848)
--(axis cs:76,9437697.53414486)
--(axis cs:75,9416903.62869848)
--(axis cs:74,9395347.03905596)
--(axis cs:73,9372999.79129081)
--(axis cs:72,9349832.88544165)
--(axis cs:71,9325816.25787908)
--(axis cs:70,9300918.74229221)
--(axis cs:69,9275108.02924421)
--(axis cs:68,9248350.62424441)
--(axis cs:67,9220611.80428267)
--(axis cs:66,9191855.57276932)
--(axis cs:65,9162044.6128226)
--(axis cs:64,9131140.23884255)
--(axis cs:63,9099102.34630889)
--(axis cs:62,9065889.35973743)
--(axis cs:61,9031458.17872773)
--(axis cs:60,8995764.12203181)
--(axis cs:59,8958760.86957141)
--(axis cs:58,8920400.40232855)
--(axis cs:57,8880632.94003143)
--(axis cs:56,8839406.87655464)
--(axis cs:55,8796668.71295009)
--(axis cs:54,8752362.98802155)
--(axis cs:53,8706432.2063528)
--(axis cs:52,8658816.76369601)
--(axis cs:51,8609454.86962346)
--(axis cs:50,8558282.46734225)
--(axis cs:49,8505233.15056802)
--(axis cs:48,8450238.07734963)
--(axis cs:47,8393225.8807332)
--(axis cs:46,8334122.57614929)
--(axis cs:45,8272851.46540334)
--(axis cs:44,8209333.03714448)
--(axis cs:43,8143484.86368368)
--(axis cs:42,8075221.49402717)
--(axis cs:41,8004454.34298625)
--(axis cs:40,7931091.57621933)
--(axis cs:39,7855037.99105654)
--(axis cs:38,7776194.89295156)
--(axis cs:37,7694459.96739886)
--(axis cs:36,7609727.14714768)
--(axis cs:35,7521886.4745359)
--(axis cs:34,7430823.95875695)
--(axis cs:33,7336421.42786)
--(axis cs:32,7238556.37526495)
--(axis cs:31,7137101.8005458)
--(axis cs:30,7031926.04419065)
--(axis cs:29,6922892.61597073)
--(axis cs:28,6809860.01642082)
--(axis cs:27,6692681.55070411)
--(axis cs:26,6571205.13373041)
--(axis cs:25,6445273.08467424)
--(axis cs:24,6314721.90774858)
--(axis cs:23,6179382.05377498)
--(axis cs:22,6039077.65293536)
--(axis cs:21,5893626.20162534)
--(axis cs:20,5742838.17291274)
--(axis cs:19,5586516.49598682)
--(axis cs:18,5424455.80662284)
--(axis cs:17,5256441.29272142)
--(axis cs:16,5082246.81879295)
--(axis cs:15,4901631.76117365)
--(axis cs:14,4714335.53246557)
--(axis cs:13,4520067.95857687)
--(axis cs:12,4318492.20598371)
--(axis cs:11,4109194.32108504)
--(axis cs:10,3891628.70385256)
--(axis cs:9,3665020.31498473)
--(axis cs:8,3428189.08950038)
--(axis cs:7,3179234.46958726)
--(axis cs:6,2914968.41015478)
--(axis cs:5,2629896.09151767)
--(axis cs:4,2314383.31733341)
--(axis cs:3,1950913.74978202)
--(axis cs:2,1501179.71974292)
--(axis cs:1,906037.635342012)
--(axis cs:0,68047.525)
--cycle;
\addplot [semithick, red]
table {%
0 67833.113
1 841159.851353783
2 1343169.58745087
3 1717270.65904687
4 2025086.09205112
5 2294714.20961858
6 2539782.69678325
7 2767324.12001381
8 2981200.30992511
9 3183671.71460484
10 3376154.07336909
11 3559600.29600263
12 3734700.77456043
13 3901991.78680445
14 4061915.63447371
15 4214854.68466948
16 4361150.98203524
17 4501117.74796801
18 4635046.2600451
19 4763210.07433734
20 4885867.70609752
21 5003264.40813163
22 5115633.41553022
23 5223196.87040701
24 5326166.55098112
25 5424744.47768114
26 5519123.43897896
27 5609487.46223333
28 5696012.24466683
29 5778865.55366884
30 5858207.60214335
31 5934191.40257879
32 6006963.10230992
33 6076662.3017223
34 6143422.35671426
35 6207370.66646387
36 6268628.94737754
37 6327313.49398495
38 6383535.42746679
39 6437400.93244459
40 6489011.48261723
41 6538464.05579193
42 6585851.33882603
43 6631261.9229679
44 6674780.49005949
45 6716487.99003989
46 6756461.81016685
47 6794775.93635275
48 6831501.10699187
49 6866704.95963737
50 6900452.17086875
51 6932804.58967404
52 6963821.36465516
53 6993559.06534983
54 7022071.79794915
55 7049411.31567651
56 7075627.12408053
57 7100766.58148254
58 7124874.99480741
59 7147995.71101559
60 7170170.20434353
61 7191438.15954995
62 7211837.55135554
63 7231404.72025488
64 7250174.44487085
65 7268180.01101334
66 7285453.2775965
67 7302024.73956142
68 7317923.58794392
69 7333177.76722059
70 7347814.03005981
71 7361857.98959837
72 7375334.16935878
73 7388266.05091655
74 7400676.11942176
75 7412585.90707427
76 7424016.03464704
77 7434986.25114775
78 7445515.47170456
79 7455621.81375763
80 7465322.63163453
81 7474634.5495836
82 7483573.49333604
83 7492154.72026414
84 7500392.84819989
85 7508301.88297517
86 7515895.24474181
87 7523185.79312721
88 7530185.85127846
89 7536907.22884554
90 7543361.24395167
91 7549558.74419684
92 7555510.12673823
93 7561225.35748927
94 7566713.98947716
95 7571985.18039677
96 7577047.70939721
97 7581909.99313539
98 7586580.10112979
99 7591065.77044548
100 7595374.41974072
101 7599513.16270341
102 7603488.82090482
103 7607307.93609651
104 7610976.78197539
105 7614501.37544038
106 7617887.48736348
107 7621140.65289671
108 7624266.18133542
109 7627269.16555777
110 7630154.49105901
111 7632926.84459843
112 7635590.72247622
113 7638150.43845626
114 7640610.13135074
115 7642973.77228116
116 7645245.17163013
117 7647427.98569739
118 7649525.72307299
119 7651541.75074009
120 7653479.29991904
121 7655341.4716641
122 7657131.24222357
123 7658851.46817353
124 7660504.8913351
125 7662094.14348456
126 7663621.75086532
127 7665090.13851028
128 7666501.63438279
129 7667858.47334401
130 7669162.80095414
131 7670416.67711473
132 7671622.07955882
133 7672780.90719543
134 7673894.98331475
135 7674966.05865982
136 7675995.81437059
137 7676985.86480565
138 7677937.76024696
139 7678852.98949245
140 7679732.98234138
141 7680579.1119769
142 7681392.69725027
143 7682175.00487082
144 7682927.25150567
145 7683650.60579305
146 7684346.19027286
147 7685015.08323789
148 7685658.32050911
149 7686276.89713825
150 7686871.7690406
151 7687443.8545611
152 7687994.03597642
153 7688523.16093575
154 7689032.04384284
155 7689521.46718173
156 7689992.18278861
157 7690444.91307195
158 7690880.35218301
159 7691299.16713902
160 7691701.99890071
161 7692089.46340628
162 7692462.15256355
163 7692820.635202
164 7693165.45798639
165 7693497.14629352
166 7693816.20505367
167 7694123.1195582
168 7694418.35623457
169 7694702.36339037
170 7694975.57192735
171 7695238.39602686
172 7695491.23380783
173 7695734.46795841
174 7695968.4663423
175 7696193.58258091
176 7696410.1566122
177 7696618.51522728
178 7696818.97258562
179 7697011.83070966
180 7697197.37995986
181 7697375.89949071
182 7697547.65768877
183 7697712.91259321
184 7697871.91229971
185 7698024.89534835
186 7698172.09109613
187 7698313.72007472
188 7698449.99433414
189 7698581.11777277
190 7698707.28645437
191 7698828.6889126
192 7698945.50644352
193 7699057.91338652
194 7699166.07739431
195 7699270.15969216
196 7699370.31532701
197 7699466.69340679
198 7699559.43733034
199 7699648.68500829
200 7699734.56907531
201 7699817.21709405
202 7699896.75175102
203 7699973.29104493
204 7700046.94846752
205 7700117.83317747
206 7700186.05016743
207 7700251.7004245
208 7700314.88108458
209 7700375.68558056
210 7700434.20378483
211 7700490.52214621
212 7700544.72382155
213 7700596.88880223
214 7700647.0940357
215 7700695.41354237
216 7700741.91852791
217 7700786.6774912
218 7700829.7563281
219 7700871.21843121
220 7700911.12478575
221 7700949.53406175
222 7700986.50270262
223 7701022.08501038
224 7701056.33322752
225 7701089.29761573
226 7701121.02653159
227 7701151.56649936
228 7701180.96228089
229 7701209.25694293
230 7701236.49192178
231 7701262.70708546
232 7701287.94079353
233 7701312.22995453
234 7701335.61008133
235 7701358.1153442
236 7701379.77862194
237 7701400.63155109
238 7701420.70457309
239 7701440.0269798
240 7701458.62695718
241 7701476.5316273
242 7701493.7670888
243 7701510.35845571
};
\end{axis}
\end{tikzpicture}
\hspace{0.9cm}
\caption{Aggregated cumulative costs and bounds.}
\label{fig:sa2}
\end{figure}

The reward $rew$ associated with each state represents the daily cost per patient. In a scale where one corresponds to one pound, we already assumed $rew(A)=100$, $rew(L)=50$, while the reward of $D$ is set to zero. Under these assumption, the corresponding MRM $(M,rew)$ is used to predict the expected annual cumulated cost of each department. This is obtained from the initial numbers $k(A)$ and $k(L)$ of patients in acute care and long stay:
\begin{equation}\label{eq:combined_costs}
cost := \sum_{s\in\{A,L\}} k(s) \cdot E[Rew_{D}^{\leq n}](s)\,.
\end{equation}
This cost can be therefore computed separately for each department. A sensitivity analysis with respect to the transition probabilities in Table \ref{tab:parameters} consists in considering the interval spanned by the extreme values of each one of these parameters and consequently define a credal transition matrix. The corresponding IMRM can be used to compute the lower and upper cumulative costs for different values of $s$, to be eventually combined as in Equation \eqref{eq:combined_costs} with the aggregated numbers about the patients of the three departments. Figure \ref{fig:sa2} show the result for a horizon of 20 years.

Finally, assume that departments are sustainable if and only if the total cumulative cost per patient until the patient is discharged or dies is less than or equal to a given threshold $r:= 15'000$ in a time horizon of one year. This corresponds to $E_{\leq 15'000} \{D\}$ in PRCTL and $\overline{E}_{\leq 15'000} \{D\}$ in IPRCTL. Following Equations \eqref{eq:exprewformula} and \eqref{eq:iecr} we can check this formula by computing the corresponding expected cumulative rewards, whose values for the different starting states are depicted in Table \ref{tab:exps}. 

\begin{table}[htp!]
\begin{center}
\begin{tabular}{crrrrr}
\toprule
$s$&\multicolumn{3}{c}{$E[Rew_{\{D\}}^{\leq 365}](s)$}&$\underline{E}$&$\overline{E}$\\
Dep.&1&2&3\\
\midrule
$A$& $5'832$  & $3'372$ & $4'009$ & $2'910$ & $6'421$ \\
$L$& $14'850$  & $14'600$ & $13'437$ & $13'437$ & $14'850$ \\
\bottomrule
\end{tabular}
\end{center}
\caption{Yearly cumulative costs for single patients.}\label{tab:exps}
\end{table}

Both formulae are satisfied, thus making the sustainability of each department robust even with respect to an imprecise evaluation of the conversion and dismissing rates. 

\section{Conclusions}\label{sec:conc}
An imprecise-probabilistic generalization of PRCTL, called IPRCTL, has been presented together with inference algorithms to compute expected cumulative rewards and bounded-reward probabilities. 
IPRCTL represent a first step toward the development of an imprecise PCTL based on \emph{imprecise Markov decision processes}. Although such processes have been already considered (e.g., \citet{delgado2011using}) their application to model checking is an open area of investigation. The same holds for \emph{imprecise continuous-time Markov chains}, that have been subject of intense research in the very last years (e.g., \citet{krak2017imprecise}) and whose application to model checking represent an open challenge we want to explore as a necessary future work.

\bibliography{main.bbl}
\end{document}